\documentclass[10pt,aps,prd,superscriptaddress,showpacs,amsmath,amssymb,twocolumn,nofootinbib]{revtex4-1}
\usepackage{physics}
\usepackage{graphicx} 
\usepackage[colorlinks=true,allcolors=blue]{hyperref}
\graphicspath{{./figures/}}
\usepackage[normalem]{ulem}
\usepackage{tikz}
\tikzset{every picture/.style={line width=0.2mm}}
\definecolor{pred}{RGB}{238,28,37}
\definecolor{pblue}{RGB}{48,49,146}
\definecolor{pgreen}{RGB}{00,163,80}
\usetikzlibrary{decorations.pathmorphing}
\tikzset{snake it/.style={decorate, decoration=snake}}

\begin{document}

\newcommand{\JLU}{Institut f\"ur Theoretische Physik,
  Justus-Liebig-Universit\"at, 
  35392 Giessen, Germany}   
\newcommand{\HFHF}{Helmholtz Research Academy Hesse for FAIR (HFHF), Campus Giessen, 35392 Giessen, Germany}
\newcommand{\MD}{Department of Physics and Maryland Center for Fundamental Physics,
University of Maryland, College Park, MD 20742 USA}
\newcommand{\IISC}{Centre for High Energy Physics, Indian Institute of Science, Bangalore, 560012, India}
\newcommand{\UB}{Fakult\"at f\"ur Physik, Universit\"at Bielefeld, D-33615 Bielefeld, Germany}

\title{Tricritical dynamics in $\boldsymbol{\mathrm{^3He}$–$\mathrm{^4He}}$ mixtures and QCD}

\author{Maneesha Pradeep}
\affiliation{\IISC}

\author{Johannes V. Roth}
\affiliation{\UB}

\author{Lorenz von Smekal}
\affiliation{\JLU}
\affiliation{\HFHF}

\date{August 2026}

\begin{abstract}
In this paper we study the universal critical dynamics near the tricritical points in $\mathrm{^3He}$--$\mathrm{^4He}$ mixtures and QCD with two massless quark flavors. In particular, we use the real-time formulation of the functional renormalization group to understand how the tricritical region connects the $O(4)$ Model~G dynamics of the two-flavor chiral phase transition to that of Model H for the QCD critical point with explicitly broken chiral symmetry. 
We show that in presence of an additional subdiffusive energy-like density, the tricritical dynamics inherits the strong dynamic scaling of Model G, where the critical fluctuations of order parameter and conserved $O(N)$ charges relax at identical rates. The strong scaling in the planar $O(2)$ case relevant for the $\lambda$-line and the tricritical point in the liquid  $\mathrm{^3He}$--$\mathrm{^4He}$ mixtures arises as an interesting limiting case. 
Including the explicit symmetry breaking by the finite  light-quark masses in QCD, the energy-like density of the tricritical dynamics mixes linearly with the $Z_2$ order parameter to eventually produce the characteristic slow fluctuations of the entropy per baryon near the QCD critical point.        
\end{abstract}
\maketitle

\section{Introduction}

Critical points are ubiquitous in nature, and the behavior of physical systems near these points exhibits universality in multiple respects. Static universality arises because critical phenomena are governed by physics at length scales much larger than any microscopic interaction length. As a result, the static universality class, which groups together all systems showing the same scaling behavior near their respective critical points in the thermodynamic limit, is {in absence of long-range interactions}
determined solely by the spatial dimension $d$ of the system and the symmetry of the order parameter. 

Near criticality, the approach to equilibrium slows down to due to the emergence of long-range correlations and the concomitant slow relaxation of thermodynamic quantities. Remarkably, this slow dynamics is itself universal and depends only on the symmetries and conservation laws of the relevant slow modes. The dynamic universality classes, originally classified by Hohenberg and Halperin \cite{Hohenberg:1977ym}, categorize systems that share identical scaling behavior in time near critical points.

In systems with rich phase diagrams, featuring multiple critical and multicritical points such as tricritical points, a single physical system can exhibit transitions between different universality classes. Such crossovers between universal regimes occur for example in systems like magnetic materials \cite{PhysRevLett.33.1098,PhysRevB.13.412,PhysRevB.19.5864} and $\mathrm{^3He}$--$\mathrm{^4He}$ mixtures \cite{PhysRevLett.24.715}, often when an exact symmetry of the system is explicitly broken. 

\begin{figure}
    \centering
    \leftline{\includegraphics[width=0.45\linewidth]{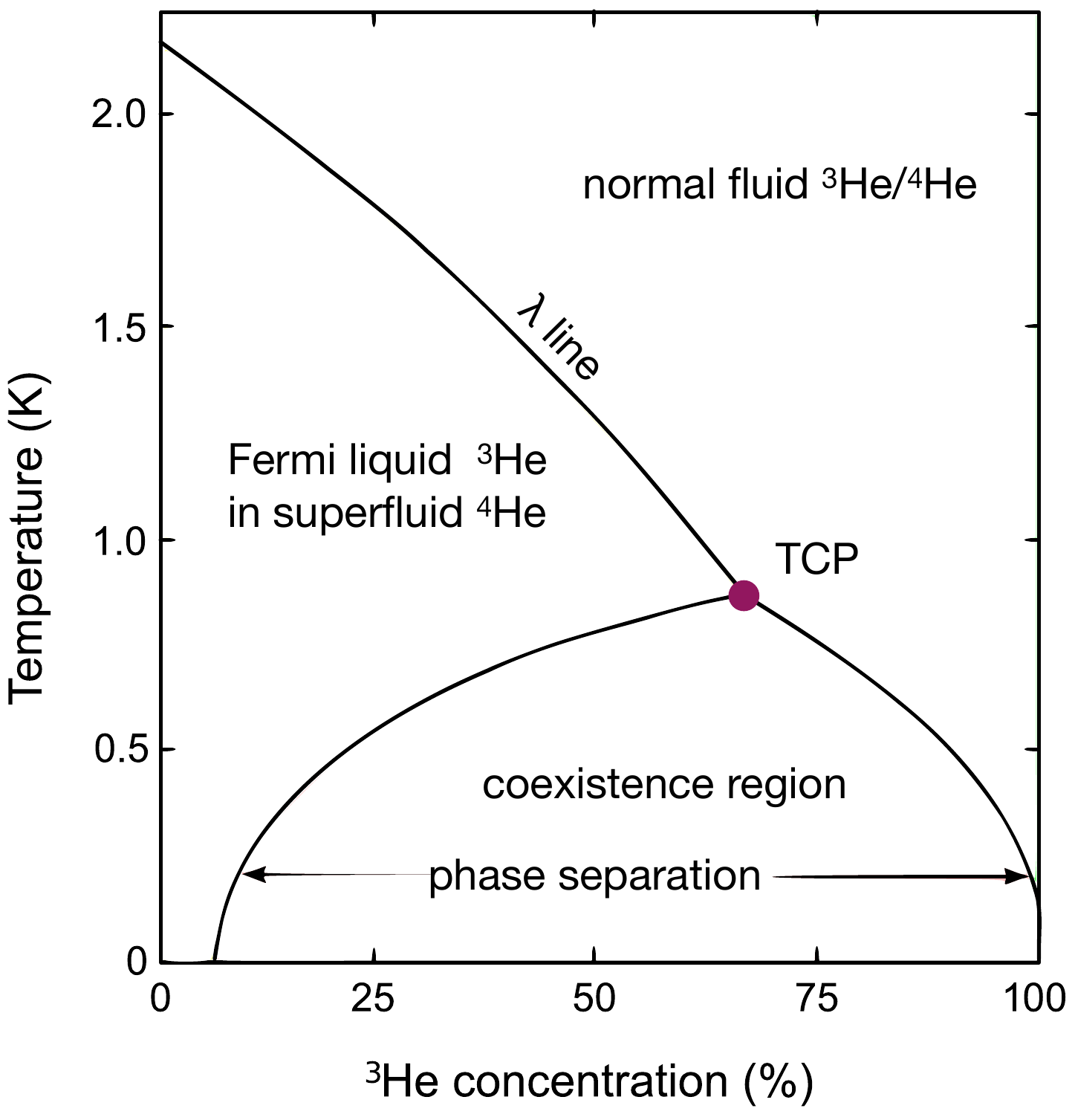}}
     \vspace{-3.9cm}
    \rightline{\includegraphics[width=0.50\linewidth]{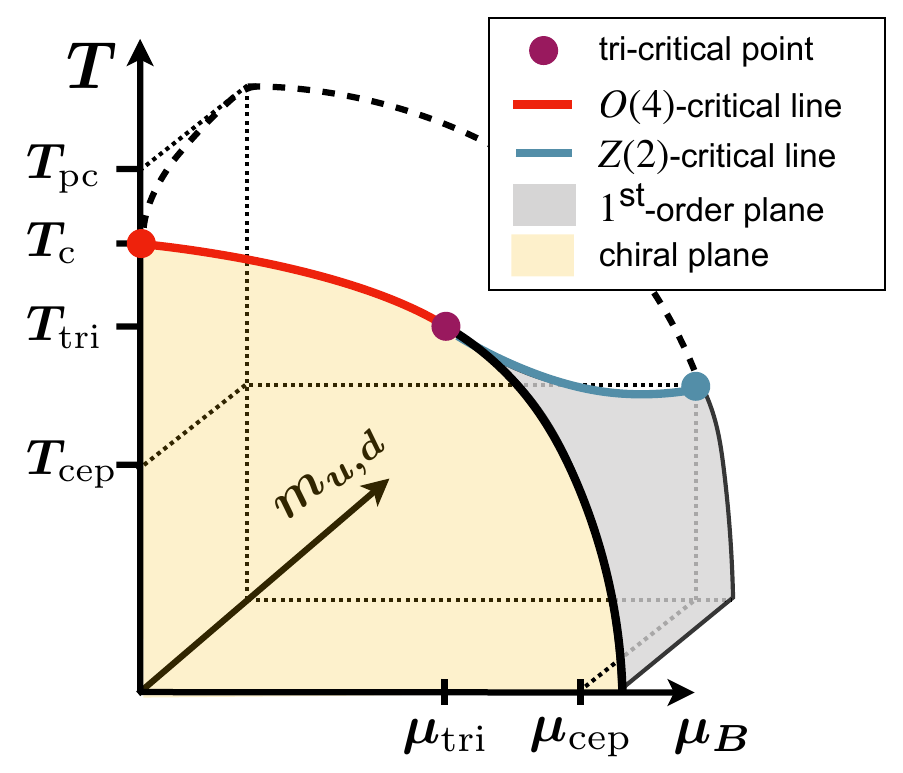}}
    \caption{Phase diagram of liquid $\mathrm{^3He}$–$\mathrm{^4He}$ mixtures at saturated vapor pressure (left), adapted from \cite{wiki:Helium_phase_diagram}, and the phase diagram of QCD in the two-flavor chiral limit (right) from \cite{Ding:2024sux}.}  
    \label{fig:pdiag5}
\end{figure}

A particularly compelling example arises in the context of the QCD phase diagram. In the two-flavor chiral limit (i.e., with massless up and down quarks), the QCD Lagrangian possesses an exact $SU(2)_L\times SU(2)_R$ chiral symmetry, which is spontaneously broken in the vacuum down to the isospin $SU(2)_V$ and restored at high temperatures. The resulting phase diagram contains regions of broken and restored chiral symmetry, separated by a phase transition. At zero baryon chemical potential, when assuming a sufficiently strong axial $U(1)_A$ anomaly, one therefore concludes that this transition is of second order in the $O(4)$ universality class \cite{Pisarski:1983ms}. As the baryon chemical potential is increased, the corresponding line of $O(4)$ critical points is then expected to terminate at a tricritical point, beyond which the transition becomes first order \cite{Halasz:1998qr}. The conjectured phase diagram is illustrated on the right-hand side of Fig.~\ref{fig:pdiag5}. This can be compared with the established phase diagram of $\mathrm{^3He}$--$\mathrm{^4He}$ mixtures, which is shown on the left-hand side. 
As a central result of this work, we shall find that both systems show similar critical dynamics near their respective tricritical points, even though the underyling static universality classes (i.e., $O(2)$ and $O(4)$) are different.

An argument by Rajagopal and Wilczek \cite{Rajagopal:1992qz} suggests that the dynamic universality class of the chiral phase transition along the $O(4)$ line is that of Model~G from the Halperin-Hohenberg classification \cite{Hohenberg:1977ym}, i.e.,~that of a Heisenberg antiferromagnet, but generalized to a four-component order parameter. 
In Model~G, the non-conserved order parameter $\phi$ relaxes slowly with a dispersion relation given by $\omega_\phi \sim p^{3/2}$ in three spatial dimensions.  Energy-density fluctuations are usually argued to be irrelevant for the corresponding critical dynamics, since the specific-heat exponent $\alpha<0$ is negative in the $O(4)$ universality class (see, e.g.,~Ref.~\cite{Nakano:2011re}). However, this argument does not apply to the tricritical point, since the tricritical specific-heat exponent $\alpha_t=1/2$ of the corresponding $\phi^6$-theory is positive \cite{Lawrie_1979}. Because of this, fluctuations of energy density might become relevant for tricritical dynamics and should be included. In this work, we study the effect of energy-density fluctuations near a tricritical point with a novel real-time formulation of the functional renormalization group (rFRG), which was already successfully used to study the non-perturbative fixed-point structure and dynamic scaling of the four-component Model~G and Model~H in earlier work \cite{Roth:2024rbi,Roth:2024hcu,Roth:2026zrs}.

When current quark masses are turned on, this tricritical point can give rise to a new line of second-order transitions in the $Z_2$ universality class. If realized in QCD, this $Z_2$ line would naturally account for the presence of a critical point in the phase diagram of QCD with physical quark masses. The critical dynamics near the $Z_2$ line is described by Model~H \cite{Son:2004iv}, where a conserved order parameter relaxes according to the dispersion relation given by $\omega_\phi \sim p^{3+x_{\eta}}$ with the small dynamic critical exponent $x_{\eta}$ of the shear-viscosity given roughly by $x_{\eta} \approx 0.05$ \cite{10.1143/PTP.55.1384}.
The reorganization of spectral modes  
that occurs when the $O(4)$ symmetry is explicitly broken by non-zero current quark masses should be similarly described by the tricritical point.
Employing a mean-field approximation, we shall recover the result of Ref.~\cite{Fujii:2004jt} that
the explicit breaking of $O(4)$ symmetry near the tricritical point leads to mixing between an energy-like density and the chiral order parameter. At long timescales, only a combination of these two remains, which becomes the critical mode of Model~H. In addition, we shall find that the pions acquire a finite pole mass 
and thus decouple from the critical dynamics on long timescales.  

This work is organized as follows. In Sec.~\ref{sec:tricritDynInQCD}, we first review the argument behind the dynamic universality class of the $O(4)$ line. We consider a generalization of Model~G to an $N$-component order parameter, which is also known as the S\'asvari-Schwabl-Sz\'epfalusy (SSS) model \cite{SASVARI1975108}. For $N=4$, the SSS model corresponds to the extension of Model~G by Rajagopal and Wilczek \cite{Rajagopal:1992qz}. We supplement the equations of the SSS model with an \mbox{energy(-like)} density, which is necessary at the tricritical point, since the specific-heat exponent $\alpha_t=1/2$ is positive there. We find that, for $N=2$, the resulting model describes tricritical dynamics in $\mathrm{^3He}$–$\mathrm{^4He}$ mixtures \cite{PhysRevLett.29.1661,GroverSwift1973,PhysRevB.15.1427,FolkMoser2007}. In Sec.~\ref{sec:rFRG}, we employ the rFRG to derive non-perturbative flow equations for the kinetic coefficients of our model as a function of $d$ and $N$. In Sec.~\ref{sec:results}, we study the fixed-point structure of flow equations in $d=3$ spatial dimensions, distinguishing between weak and strong dynamic scaling regimes, depending on the value of $N$. In Sec.~\ref{sec:explSymmBreak}, we study the transition to the $Z_2$ line by including an external symmetry-breaking field in a mean-field approximation. In Sec.~\ref{sec:summaryAndOutlook}, we summarize our results and give a brief outlook for future research opportunities. Several appendices are added to provide additional technical details, which we have omitted in the main text.

\section{Tricritical dynamics in QCD}
\label{sec:tricritDynInQCD}

At small baryon chemical potential $\mu_B$, the two-flavor chiral phase transition is conjectured to belong to the dynamic universality class of Model~G from the Halperin-Hohenberg classification \cite{Hohenberg:1977ym} (which describes the critical dynamics of a Heisenberg antiferromagnet) generalized to a four-component order parameter \cite{Rajagopal:1992qz}. This can be understood as follows \cite{Son:2001ff}. 
Spontaneous breaking of chiral symmetry is signaled by a non-vanishing  chiral condensate  
$\Phi_{ij} = \langle \bar{q}_{L i} q_{R j} \rangle$, 
where $q_i$ with $i=1,2$ enumerates the two quark flavors. 
Under chiral transformations $U,V \in SU(2)$, the matrix $\Phi$ behaves as
\begin{equation}
    \Phi \to U^\dagger \Phi V \,. \label{eq:PhiChiralTransform}
\end{equation}
With the axial $U(1)_A$ anomalously broken,
states on the $S^3$ vacuum manifold  can be parameterized by the phase $\Sigma = \vec{\varphi}\cdot\vec{\tau} \in su(2)$ of the chiral condensate, $\Phi \equiv \sqrt{\rho}\, e^{i\Sigma}$, where $\tau_i$ denote the Pauli matrices.
Long-wavelength modulations of $\Sigma$
relax arbitrarily slowly, 
since a non-zero but spatially constant phase would simply describe a different vacuum state and would thus not dissipate in time. As such, $\Sigma$ enters the hydrodynamics of the system.
Moreover, the Partially Conserved Axial Current (PCAC) relation entails that for vanishing current quark masses
the iso-axial-vector charge densities $A_i = \langle \tfrac{1}{2}\bar{q}\gamma^0 \gamma^5 \tau_i q \rangle$ satisfy exact conservation laws, in addition to the iso-vector charge densities $V_i = \langle \tfrac{1}{2}\bar{q}\gamma^0 \tau_i q \rangle$.
The mixing of $\varphi_i$ and $A_i$ below the critical temperature 
gives rise to propagating modes with characteristic frequency $\omega = vp$ \cite{Son:2002ci}, which
correspond to soft pions. As shown in \cite{Son:2002ci}, the pion velocity $v$ is given by $v = f/\sqrt{\chi_{n}}$, where $f$ denotes the pion decay constant, and $\chi_{n}$ the susceptibility of the iso-axial-vector charge, i.e.,~$v$ is fully determined by static quantities. 
(This is akin to the second-sound mode in superfluid ${}^4\mathrm{He}$.) 
As the critical temperature is approached from below,
static critical scaling entails that $f$ vanishes as $f\sim (T_c-T)^{\nu(d/2-1)} \sim \xi^{-d/2+1}$, while $\chi_n$ stays finite \cite{Son:2001ff}. Right at the boundary of applicability of the linearized hydrodynamic theory $p\sim\xi^{-1}$ one obtains the characteristic frequency of soft pions in $d=3$ dimensions,
\begin{equation}
    \omega \sim \xi^{-3/2} \,. \label{eq:piCharFreq}
\end{equation}     

Choosing the vacuum to conserve parity and hence be aligned with the identity matrix, $\Phi \to \sigma\mathbf{1}$, we can 
write $\Phi =  
\sigma+i\vec{\pi}\cdot\vec{\tau}$, where
the chiral transformations \eqref{eq:PhiChiralTransform} correspond to
$O(4)$ transformations of $\phi = (\sigma,\vec{\pi})^T$.
When chiral symmetry is restored, the correlators of $\sigma$ and $\vec{\pi}$ become degenerate, and $\sigma$ inherits the characteristic frequency \eqref{eq:piCharFreq}. Comparing with the expectation $\omega_{\phi} \sim \xi^{-z_{\phi}}$ for the order parameter from critical slowing down near a continuous phase transition, one obtains  $z_{\phi} =3/2$, i.e., the dynamic critical exponent of Model~G. 

The specific-heat exponent $\alpha<0$ is negative in $O(4)$, with negative amplitudes of the singular part, which means that the isochoric specific heat $c_v = \partial\varepsilon/\partial T\rvert_{V} \sim |T-T_c|^{-\alpha}$ is expected to have an upward cusp at $T_c$.\footnote{At physical quark masses $m_q$, lattice-QCD simulations indeed show a pronounced peak in the temperature derivative of the trace anomaly, which is the dominant singular contribution from the chiral phase transition at $m_q=0$ to the specific heat \cite{HotQCD:2014kol}.} In particular, this is reminiscent of the $\lambda$-transition in ${}^4\mathrm{He}$, where the specific heat shows the characteristic $\lambda$-like shape precisely because $\alpha<0$ 
in the $O(2)$ universality class as well \cite{PhysRevB.68.174518}, in fact it is negative in all $O(N)$ universality classes with $N\ge 2$ in $d=3$ dimensions. As a consequence, the static fluctuations of the energy density $V\langle(\delta \varepsilon)^2\rangle = T^2 c_v$ stay finite and have no influence on the critical dynamics of the order parameter (see, e.g.,~Ref.~\cite{Nakano:2011re}). At the tricritical point, on the other hand, this argument must be revisited, since there, the (tricritical) specific-heat exponent $\alpha_t=1/2$ is positive \cite{Lawrie_1979}. In this case, $\alpha_t$ determines the divergence of the specific heat at constant pressure $c_p$, instead of $c_v$. This can be derived quite generally from the scaling hypothesis near a tricritical point (see Appendix~\ref{App:cv}), and is also observed in $\mathrm{^3He}$--$\mathrm{^4He}$ mixtures~\cite{Takada1980}. By the fluctuation-response theorem in local thermal equilibrium, $c_p = Tn_B \partial\hat{s}/\partial T\rvert_{p}$ determines the variance of the entropy per baryon $\hat{s}\equiv s/n_{B}$, which gives $V\langle (\delta\hat{s})^2\rangle = c_p/n_B^2$. Hydrodynamically, the fluctuations of $\hat{s}$ are special because they do not mix with the propagating sound modes and thus relax purely via diffusion \cite{Lifshitz:1980,Kovtun:2012rj}. 
It follows that, after the pressure has equilibrated locally through sound waves, the densities $\varepsilon$ and $n_B$ trace the diffusion of $\hat{s}$. 
Similar to Ref.~\cite{Fujii:2004jt}, we use the term `energy-like' density to denote any density that has non-zero overlap with $\delta\hat{s}$, inheriting both its static fluctuations and its diffusive dynamics on long timescales.

We therefore consider the $O(4)$ order parameter field  $\phi$ coupled to the iso\--\mbox{(axial-)}\-vector charge densities $V_i$ and $A_i$ \emph{and} to an energy-like density $\varepsilon$. It is convenient to combine
the former into an anti-symmetric tensor $n_{ab}$, setting $n_{0i} = A_i$ and $n_{ij} = \varepsilon_{ijk}V_k$. 
To study the static critical behavior, this leads to a Landau-Ginzburg-Wilson (LGW) functional as an appropriate starting point which is a variation of the form in \cite{Fujii:2004jt}, 
\begin{align}
    F &= \int d^dx \, \bigg\{ \frac{1}{2}(\vec{\nabla}\phi)^2 + \frac{m^2}{2}\phi^2 + \frac{\lambda}{4!N} (\phi^2)^2 + \frac{\kappa}{6!N^2} (\phi^2)^3 \nonumber \\
    &\hspace{2.0cm}+\frac{\iota}{2\sqrt{N}} \phi^2 \varepsilon + \frac{\varepsilon^2}{2\chi_{\varepsilon}} + \frac{n_{ab} n_{ab}}{4\chi_n} \bigg\} \, . \label{eq:freeEnergy}
\end{align}
Here, with $\phi^2 \equiv \phi_a \phi_a$, 
we have rescaled $\phi$ such that the coefficient of $(\vec{\nabla}\phi)^2$ is fixed to 1/2. The particular $(\phi^2)^3$ coupling $\kappa$ is needed for stability at the tricritical point.
Compared to Ref.~\cite{Fujii:2004jt}, we have also added a  term quadratic in $n_{ab}$, representing the non-critical equilibrium fluctuations of the charge densities with susceptibility $\chi_n$ \cite{Son:2001ff}.
Fluctuations are distributed according to $e^{-F/T}$. As discussed above, exact chiral symmetry forbids a linear mixing of $\phi$ and $\varepsilon$. 

To write down the equations of motion for the long-time dynamics near the tricritical point, we employ the rather elegant Poisson-bracket technique \cite{DZYALOSHINSKII198067}. In this approach, the Poisson brackets arise via the correspondence principle from the commutators in the microscopic theory,
\begin{align}
    \{ \phi_a, n_{bc} \} &=\delta_{ac}\phi_b-\delta_{ab}\phi_c \,, \\
    \{ n_{ab}, n_{cd} \} &=\delta_{ac}n_{bd}+\delta_{bd}n_{ac} -\delta_{ad}n_{bc}  -\delta_{bc}n_{ad} \,, \nonumber 
\end{align}
reflecting the fact that, by Noether's theorem, the charges $n_{ab}$ are the generators of the $O(4)$ transformations. The ideal equations of motion including the corresponding reversible mode couplings thus read
\begin{align}
    \frac{\partial \phi_a}{\partial t} &= g\{\phi_{a},F\} = \frac{g}{2} \{ \phi_a, n_{bc} \} \frac{\delta F}{\delta n_{bc}}  \,,  \\ \nonumber
    \frac{\partial n_{ab}}{\partial t} &= g\{n_{ab},F\} = g\{n_{ab},\phi_c\} \frac{\delta F}{\delta \phi_c} + \frac{g}{2} \{ n_{ab},n_{cd} \}\frac{\delta F}{\delta n_{cd}} \,, \\ \nonumber 
    \frac{\partial \varepsilon}{\partial t} &= g\{\varepsilon,F\} = 0 \,,
\end{align}
which ensure $dF/dt = 0$.
Introducing dissipation and noise to ensure the approach to the equilibrium distribution $e^{-F/T}$ while respecting the conservation laws, one then obtains
\begin{align} \frac{\partial \phi_a}{\partial t} &= - \Gamma^{\phi} \frac{\delta F}{\delta \phi_a} + \frac{g}{2} \{ \phi_a, n_{bc} \} \frac{\delta F}{\delta n_{bc}}  + \theta_a \,, \label{eq:eoms}  \\ \nonumber
    \frac{\partial n_{ab}}{\partial t} &= \gamma \vec{\nabla}^2 \frac{\delta F}{\delta n_{ab}} + g\{n_{ab},\phi_c\} \frac{\delta F}{\delta \phi_c} \;+ \\ \nonumber
    &\hspace{3.0cm} \frac{g}{2} \{ n_{ab},n_{cd} \}\frac{\delta F}{\delta n_{cd}} + \zeta_{ab} \,, \\ \nonumber 
    \frac{\partial \varepsilon}{\partial t} &= \mu \vec{\nabla}^2 \frac{\delta F}{\delta \varepsilon} + \Xi \,.
\end{align}
$\Gamma^{\phi}$ denotes the order-parameter damping rate, $\gamma$ the iso-(axial-)vector charge mobility, and $\mu$ is proportional to the heat conductivity. 
The Langevin noise terms $\theta_a$, $\zeta_{ab}$, and $\Xi$ have vanishing expectation values, and variances set by fluctuation-dissipation relations, 
\begin{align}
    \langle \theta_a(t,\vec{x})\theta_b(t',\vec{x}') \rangle &= 2 \Gamma^{\phi} T \delta_{ab} \delta(\vec{x}-\vec{x}') \delta(t-t') \,, \label{eq:noiseVariances} \\ \nonumber
    \langle \zeta_{ab}(t,\vec{x}) \zeta_{cd}(t',\vec{x}') \rangle &= \\ \nonumber
    &\hspace{-1.5cm} -2\gamma T (\delta_{ac}\delta_{bd}-\delta_{ad}\delta_{bc}) \vec{\nabla}^2  \delta(\vec{x}-\vec{x}') \delta(t-t')  \,, \\ \nonumber
    \langle \Xi(t,\vec{x})\Xi(t',\vec{x}')\rangle & =  -2 \mu T \, \vec{\nabla}^2 \delta(\vec{x}-\vec{x}') \delta(t-t') \, .
\end{align}

Our model \eqref{eq:eoms} can be trivially generalized to an $N$-component order parameter, in which case it would correspond to the SSS model with an additional conserved scalar density statically coupled to the order parameter. 

In particular, in the $N=2$ case our model \eqref{eq:eoms} leads to the same universal critical dynamics as the one by Siggia and Nelson \cite{PhysRevB.15.1427}, which aims to describe tricritical dynamics in mixtures of superfluid $\mathrm{^4He}$ and $\mathrm{^3He}$. In the nomenclature of Ref.~\cite{Folk_2006} this model is called Model~F', to emphasize that it extends the original Model~F by Halperin and Hohenberg \cite{Hohenberg:1977ym} to mixtures with ${}^3\mathrm{He}$. In its original form, Model~F describes the critical dynamics of ${}^4\mathrm{He}$ near the $\lambda$-transition. More specifically, it describes a non-conserved $U(1)$ order parameter (usually regarded as the wave function of the Bose-Einstein condensate), coupled both statically and dynamically to the fluctuations of the entropy per particle. Model~F', on the other hand, involves the concentration of ${}^3\mathrm{He}$ as an additional conserved density \cite{Folk_2006}. Applying the naming conventions of Ref.~\cite{Folk_2006}, one could refer to our model \eqref{eq:eoms} for general $N$ as the \emph{SSS' model}, since it supplements the SSS model by an additional conserved (scalar) energy-like density.

For $N=2$,
an obvious difference of our model, as defined by Eqs.~\eqref{eq:eoms} (without the non-Abelian mode couplings $\propto \{ n_{ab}, n_{cd} \}$), compared to Model F' is that the latter allows the damping rate of the $U(1)$ order parameter to be complex. While the corresponding imaginary part certainly affects the non-universal dynamics away from the $\lambda$-line, renormalization-group studies in $d=4-\epsilon$ dimensions indicate it to be irrelevant for the asymptotic critical behavior in $d=3$ dimensions \cite{PhysRevB.15.1427,Onuki1983}. Finally, we mention that Model~F' is known to be problematic at the tricritical point when $\mathcal{O}(\epsilon^2)$ corrections are taken into account  \cite{FolkMoser2007}. The rFRG, on the other hand, is inherently non-perturbative, so one may hope to gain novel insights into tricritical dynamics in Model~F'. 

\section{Real-time functional renormalization group}
\label{sec:rFRG}

\subsection{Statics}
\label{sec:rFRGA}

To systematically incorporate the effects of critical fluctuations near a second-order phase transition, we use the functional renormalization group (FRG) in the formulation pioneered by Wetterich \cite{Wetterich:1992yh}. The idea is to artificially suppress fluctuations of modes with wavenumbers $p \lesssim k$ by adding an infrared (IR) regulator $R_k^{\phi}(p)$ to the free energy \eqref{eq:freeEnergy}, and then study the dependence of the resulting coarse-grained free energy $F_k = F_k[\phi,\varepsilon,n]$ on the FRG scale $k$, which is described by the exact flow equation 
\begin{align}
    \partial_k F_k =\frac{T}{2} \Tr\left\{ \partial_k R_k\big(F_k^{(2)}+R_k\big)^{-1} \right\} \, ,
    \label{eq:staticFlow}
\end{align}
where $F_k^{(2)}$ is the matrix of second functional derivatives of $F$,
\begin{align}
    F_k^{(2)} = \begin{pmatrix}
        F_k^{\phi\phi} & F_k^{\phi\varepsilon} & F_k^{\phi n} \\
        F_k^{\varepsilon\phi} & F_k^{\varepsilon\varepsilon}  & F_k^{\varepsilon n} \\
        F_k^{n\phi}  & F_k^{n\varepsilon} & F_k^{n n}
    \end{pmatrix} \, ,
\end{align}
in which we have used a shorthand notation for functional derivatives, e.g.,
\begin{align}
    F_k^{\phi_a\varepsilon}(\vec{x},\vec{y}) \equiv \frac{\delta^2 F_k}{\delta \phi_a(\vec{x})\delta\varepsilon(\vec{y})} \,, \quad\text{etc.} \,.
\end{align}
In particular, at some sufficiently large UV cutoff $k=\Lambda$ the LGW functional \eqref{eq:freeEnergy} is recovered, while for $k=0$ all fluctuations are included, such that $F_{k=0}$ equals the full free energy in the presence of inhomogeneous field expectation values.

Throughout this work, we choose the optimized regulator for the order parameter \cite{Litim:2001up},
\begin{align}
    R_k^{\phi}(p) = (k^2 - p^2)\theta(k^2 - p^2) \,. \label{eq:RPhi}
\end{align}
We do not introduce regulators for the energy-like density and the $O(N)$ charge densities. This is because the static fluctuations of $n_{ab}$ are non-critical, and thus do not need to be regulated. On the other hand, the fluctuations of the auxiliary field $\varepsilon$ are related to $\chi_{\varepsilon}$, which is divergent for $\alpha > 0$, but this divergence only builds up gradually for $k\to 0$ since it is induced by the fluctuations of $\phi$ which are already regulated due to the presence of $R_k^{\phi}(p)$.

As the basis for our truncation of the FRG flow, we start from a coarse-grained free energy $F_k$ in local potential approximation (LPA) given by
\begin{align}
    F_k = \int d^d x \bigg\{ \frac{1}{2}(\vec{\nabla}\phi)^2 + U_k(\rho,\varepsilon) + \frac{n_{ab}n_{ab}}{4\chi_n} \bigg\} \, , \label{eq:freeEnergyLPA}
\end{align}
with field invariant $\rho \equiv \phi_a\phi_a$, the scale-dependent effective potential $U_k(\rho,\varepsilon)$, and the non-critical $O(N)$ charge densities $n_{ab}$. In the following, we drop the charge densities $n_{ab}$ for brevity, since their fluctuations are Gaussian, and decoupled from the fluctuations of $\rho$ and $\epsilon$, and do thus not enter the FRG flow of the effective potential. In the remaining $(\phi,\varepsilon)$ field space, the Hessian $F_k^{(2)}$ of the coarse-grained free energy \eqref{eq:freeEnergyLPA} for homogeneous field expectation values $\phi=(\sigma,\vec{0})^T$ and $\varepsilon$ is given by 
\begin{align}
    &F_k^{(2)} = \\ \nonumber
    &\begin{pmatrix}
        2U' + 4\sigma^2 U'' -\vec{\nabla}^2 & 0  & \cdots &  0 & 2\sigma \dot{U}' \\
        0  & 2U' -\vec{\nabla}^2 &  &  & 0  \\
        \vdots &  & \ddots  &   & \vdots  \\
        0  &  &  &  2U'-\vec{\nabla}^2 & 0  \\
        2\sigma \dot{U}' & 0  & \cdots & 0 & \ddot{U}
    \end{pmatrix}
\end{align}
where the different field derivatives of the effective potential are denoted by $U_k' \equiv \partial U/\partial \rho$ and $\dot{U}_k \equiv \partial U/\partial \varepsilon$. 
Inverting $F_k^{(2)}+R_k$ yields the static propagators of $\sigma$, $\vec{\pi}$ and $\varepsilon$,
\begin{align}
    G_k^{\sigma\sigma}(\vec{p}) &= \frac{1}{2U' + 4\sigma^2 (U''-\dot{U}'^2/\ddot{U})+\vec{p}^2 + R_k^{\phi}(p)}\, , \label{eq:staticResponseFncSigma} \\
    G_k^{\pi_i\pi_j}(\vec{p})  &= \frac{\delta_{ij}}{2U'+\vec{p}^2 + R_k^{\phi}(p)} \, , \\
   G_k^{\varepsilon\varepsilon}(\vec{p}) &= \frac{(2U' + 4\sigma^2 U'')}{[2U' + 4\sigma^2 (U''-\dot{U}'^2/\ddot{U})]\ddot{U}} \, ,
\end{align}
from which we read off the (squared) screening masses of the sigma and the pions,
\begin{align}
    m_{\sigma}^2 \equiv 2U' + 4\sigma^2 (U''-\dot{U}'^2/\ddot{U}) \,, \quad m_{\pi}^2 \equiv 2U' \,.
\end{align}
In particular, the screening mass $m_{\sigma}$ determines the correlation length $\xi = 1/m_{\sigma}$ of the (chiral) order parameter. 
Inserting the homogeneous field expectation values into the flow equation \eqref{eq:staticFlow} and using the optimized regulator \eqref{eq:RPhi} yields the flow of the effective potential,
\begin{align}
    \partial_k U_k &= K_d k^{d+1} T \bigg( \frac{1}{2U_k' + 4\rho( U_k'' - \dot{U}_k'^2/\ddot{U}_k )+k^2}  \label{eq:flowEffPot} \\ \nonumber 
    &\hspace{5.5cm} + \frac{N-1}{2U_k'+k^2} \bigg)
\end{align}
with $K_d \equiv \Omega_d/(2\pi)^d$, where $\Omega_d = 2\pi^{d/2}/(\Gamma(d/2)\,d)$ denotes the volume of the $d$-dimensional unit ball. The first term on the right-hand side represents the contribution from the sigma mode, while the second term represents the contribution from the $N-1$ Goldstone (pion)  modes. The extra term ($-4\rho \dot{U}_k'^2/\ddot{U}_k$) in the inverse propagator of the sigma mode is due to the mixing of the sigma mode with the energy-like density $\varepsilon$  at non-zero field expectation values. 

We consider a truncation of the full effective potential $U_k(\rho,\varepsilon)$ which is motivated by promoting the couplings in \eqref{eq:freeEnergy} to depend on the FRG scale $k$.
I.e.,~completing the square with $\lambda'_k \equiv \lambda_k - 3\iota_k^2 \chi_{\varepsilon,k}$ in \eqref{eq:freeEnergy}, our ansatz reads 
\begin{align}
    U_k(\rho,\varepsilon) &= \frac{m_k^2}{2}\rho + \frac{\lambda'_k}{4!N} \rho^2 + \frac{\kappa_k}{6!N^2} \rho^3 \; + \label{eq:effPotTrunc} \\ \nonumber
    &\hspace{2.0cm} \frac{1}{2\chi_{\varepsilon}} \left( \varepsilon - \varepsilon_{0,k} + \frac{\iota_{k} \chi_{\varepsilon,k}}{2\sqrt{N}} \rho \right)^2 \,,
\end{align}
where we identify the $\rho$-dependent minimum of $\varepsilon$,
\begin{align}
    \varepsilon_{\text{min},k}(\rho) = \varepsilon_{0,k} - \frac{\iota_{k} \chi_{\varepsilon,k}}{2\sqrt{N}}\rho \, ,
\end{align}
which satisfies $\partial U_k/\partial\varepsilon\rvert_{\varepsilon_{\text{min},k}(\rho)} = 0$.
Note that we have introduced a non-vanishing expectation value $\varepsilon_{0,k} \neq 0$ of the energy-like density, which is absent in the bare potential in \eqref{eq:freeEnergy}, but will be generated by the flow. This can be seen by recalling that $\varepsilon$ 
is essentially the expectation value of $\phi^2$, which does not factorize as soon as fluctuations of $\phi$ are included.

Now we turn to the flow equations. We project the FRG flow onto the couplings appearing in \eqref{eq:effPotTrunc} according to Appendix~\ref{sec:exactSolEffPotGaussFP}.  In particular, we find that 
$m_k^2$, $\lambda'_k$, and $\kappa_k$ satisfy the well-known LPA flow equations of the $O(N)$ scalar field theory with an effective potential expanded up to third order in $\rho$, 
\begin{align}
    \partial_k m_k^2 &= - 
    \frac{K_d k^{d+1} T (N+2) \lambda'_k}{3N (m_k^2+k^2)^2}  \label{eq:m2LambdaKappaFlow} \, , \\ \nonumber
    \partial_k \lambda'_k &= K_d k^{d+1} T \left\{
    \frac{2 (N+8) \lambda_k'^2}{3N (m_k^2+k^2)^3} -
    \frac{(N+4)\kappa_k}{5N (m_k^2+k^2)^2}  \right\} \, , \\ \nonumber
    \partial_k \kappa_k &= K_d k^{d+1} T\left\{
    \frac{2 (N+14) 
    \lambda'_k \kappa_k}{N(m_k^2+k^2)^3} - 
    \frac{10(N+26) \lambda_k'^3}{3N(m_k^2+k^2)^4} \right\} \, ,
\end{align}
and are thus independent of the coupling to $\varepsilon$. On the other hand, the flow equations for the coupling $\iota_k$ of $\phi^2 $ and the energy-like density $\varepsilon $, and the susceptibility $\chi_{\varepsilon,k}$ of the latter are given by
\begin{align}
    \partial_k \iota_{k} &= \frac{2K_d k^{d+1} T \iota_k [ \iota_{k}^2 \chi_{\varepsilon,k}+ (N+2) \lambda'_k/3N ]}{(m_k^2+k^2)^3} \,, \label{eq:iotaFlow} \\
    \partial_k \chi_{\varepsilon,k} &= -\frac{2K_d k^{d+1} T \iota_{k}^2 \chi_{\varepsilon,k}^2}{(m_k^2+k^2)^3} \,. \label{eq:chiFlow}
\end{align}

\subsection{Dynamics}

\newcommand{\Dvec}[1]{\underline{#1}}

We use the rFRG framework developed in Refs.~\cite{Roth:2024rbi,Roth:2024hcu} to derive non-perturbative flow equations for the kinetic coefficients. In particular, this framework guarantees that the flow equations in \eqref{eq:m2LambdaKappaFlow}, \eqref{eq:iotaFlow}, and \eqref{eq:chiFlow} 
are unchanged, 
and that the mode-coupling constant $g$ is protected from loop corrections, i.e.~from renormalization, by symmetry \cite{Roth:2024rbi}. 
The starting point is the Martin-Siggia-Rose (MSR) path-integral reformulation of the Langevin equations \eqref{eq:eoms}, in which the partition function is given by 
\begin{align}
    Z = \int \mathcal{D}\phi \mathcal{D}\tilde{\phi} \mathcal{D}n \mathcal{D}\tilde{n} \mathcal{D}\varepsilon \mathcal{D}\tilde{\varepsilon}\,e^{iS} = 1 \, , \label{eq:partFnc}
\end{align}
with MSR action
\begin{widetext}
\begin{align}
    S = \int_{\Dvec{x}} \bigg\{ &-\tilde{\phi}_a\left(\frac{\partial \phi_a}{\partial t} + \Gamma^{\phi} \frac{\delta F}{\delta \phi_a} - \frac{g}{2} \{ \phi_a, n_{bc} \} \frac{\delta F}{\delta n_{bc}} \right) \nonumber \\
    &-\frac{1}{2}\tilde{n}_{ab} \left( \frac{\partial n_{ab}}{\partial t} - \gamma \vec{\nabla}^2 \frac{\delta F}{\delta n_{ab}} - g\{n_{ab},\phi_c\} \frac{\delta F}{\delta \phi_c} - \frac{g}{2} \{ n_{ab},n_{cd} \}\frac{\delta F}{\delta n_{cd}} \right) \nonumber \\
    &-\tilde{\varepsilon} \left( \frac{\partial \varepsilon}{\partial t} - \mu \vec{\nabla}^2 \frac{\delta F}{\delta \varepsilon} \right)  + iT \Gamma^{\phi} \tilde{\phi}_a\tilde{\phi}_a - \frac{i}{2} T  \tilde{n}_{ab} \gamma \vec{\nabla}^2 \tilde{n}_{ab} - i T  \tilde{\varepsilon} \mu \vec{\nabla}^2 \tilde{\varepsilon} \bigg\} \, . \label{eq:MSRAction}
\end{align}
\end{widetext}
where we have introduced the compact notation $\Dvec{x} \equiv (t,\vec{x})^T$ for four-vectors (or, more generally, $(d+1)$-vectors), with a corresponding shorthand notation for integrals,
\begin{equation*}
    \int_{\Dvec{x}} \equiv \int dt \int d^dx \,. 
\end{equation*}

Introducing physical external sources $H$, $\mathcal{H}$ and $B$ for $\phi$, $n$ and $\varepsilon$ 
into the LGW functional $F$ in \eqref{eq:freeEnergy}, and `unphysical' external sources $\tilde{H}$,$\tilde{\mathcal{H}}$ and $\tilde{B}$ for $\phi$, $n$ and $\varepsilon$ on the level of the MSR action \eqref{eq:MSRAction}, we obtain a generating functional for correlation functions of the form
\begin{align}
    &Z[H,\tilde{H},\mathcal{H},\tilde{\mathcal{H}},B,\tilde{B}] = \int \mathcal{D}\phi\,\mathcal{D}\tilde{\phi}\,\mathcal{D}n\,\mathcal{D}\tilde{n}\,\mathcal{D}\varepsilon\,\mathcal{D}\tilde{\varepsilon}\, \label{eq:genFunc} \\ \nonumber
    &\hspace{0.5cm} \exp\bigg\{ iS + i\int_{\Dvec{x}} \big( \tilde{H}_a \phi_a + \frac{1}{2}\tilde{\mathcal{H}}_{ab} n_{ab} + \tilde{B}\varepsilon \big)  \\ \nonumber
    &\hspace{1.5cm} + i\int_{\Dvec{x}} H_a\big( \Gamma^{\phi} \tilde{\phi}_a + \frac{g}{2}\{\phi_a,n_{bc}\} \tilde{n}_{ab} \big) \\ \nonumber
    &\hspace{1.5cm}  + \frac{i}{2}\int_{\Dvec{x}}  \mathcal{H}_{ab} \big(-\gamma \vec{\nabla}^2 \tilde{n}_{ab} + g\{n_{ab},\phi_c\} \tilde{\phi}_c   \\ \nonumber
    &\hspace{2.0cm} + \frac{g}{2} \{n_{ab},n_{cd}\} \tilde{n}_{cd} \big) - i\int_{\Dvec{x}} B\mu\vec{\nabla}^2\tilde{\varepsilon} \bigg\} \, . 
\end{align}
In particular, the physical external sources then couple to the following composite fields,
\begin{align}
    \tilde{\Phi}_a &\equiv \Gamma^{\phi} \tilde{\phi}_a + \frac{g}{2}\{\phi_a,n_{bc}\} \tilde{n}_{ab}   \label{eq:ResponseFields} \\ \nonumber
    \tilde{N}_{ab} &\equiv -\gamma \vec{\nabla}^2 \tilde{n}_{ab} + g\{n_{ab},\phi_c\} \tilde{\phi}_c  + \frac{g}{2} \{n_{ab},n_{cd}\} \tilde{n}_{cd}  \\ \nonumber
    \tilde{\mathcal{E}} &\equiv -\mu\vec{\nabla}^2 \tilde{\varepsilon}
\end{align} 
Their relation to the standard response fields is best emphasized in matrix notation,
\begin{align}
    &\begin{pmatrix}
        \tilde{\Phi}_a \\ \tilde{N}_{bc} \\ \tilde{\mathcal{E}}
    \end{pmatrix}= \label{eq:Jacobian} \\ \nonumber 
    &\begin{pmatrix}
        \Gamma^{\phi} \delta_{ad} & \frac{g}{2}\{\phi_a,n_{ef}\} & 0 \\
        g\{n_{bc},\phi_d\}  & -\gamma \vec{\nabla}^2 \delta_{be}\delta_{cf} + \frac{g}{2} \{n_{bc},n_{ef}\} & 0 \\
        0 & 0 & -\mu\vec{\nabla}^2 
    \end{pmatrix} \!\!
    \begin{pmatrix}
        \tilde{\phi}_d \\ \tilde{n}_{ef} \\ \tilde{\varepsilon}
    \end{pmatrix} 
\end{align}
Note that the matrix only depends on the classical fields, with $\{\phi,n\}\sim \phi$ and $\{ n,n\} \sim n$. 
Using the compact superfield notation $\psi=(\phi,n,\varepsilon)^T$, $\tilde{\Psi}=(\tilde{\Phi},\tilde{N},\tilde{\mathcal{E}})^T$ and $\tilde{\psi}=(\tilde{\phi},\tilde{n},\tilde{\varepsilon})^T$, this non-linear field transformation can be compactly written as\footnote{Contraction over superfield indices is defined by
\[
    X_{A} Y_{A} \equiv X_{\phi_a} Y_{\phi_a} + \tfrac{1}{2} X_{n_{ab}} Y_{n_{ab}} + X_{\varepsilon} Y_{\varepsilon} \,.
\]}
\begin{equation}
    \tilde{\Psi}_A = \mathcal{J}_{AB}(\psi)\tilde{\psi}_B \,.
\end{equation}
Introducing `supersources' $J=(H,\mathcal{H},B)^T$ and $\tilde{J}=(\tilde{H},\tilde{\mathcal{H}},\tilde{B})^T$, the generating functional \eqref{eq:genFunc} can be compactly written as 
\begin{align}
    Z[J,\tilde{J}] &= \int \mathcal{D}\psi\,\mathcal{D}\tilde{\psi}\,\exp\big\{ iS \;+ \label{eq:genFuncSuperfield} \\ \nonumber
    &\hspace{1.0cm} i\int_{\Dvec{x}} \big( \tilde{J}_A(\Dvec{x}) \psi_A(\Dvec{x}) + J_A(\Dvec{x}) \tilde{\Psi}_A(\Dvec{x}) \big) \big\}
\end{align}
Its logarithm is the generator of connected correlation functions,
\begin{equation}
    W[J,\tilde{J}] = -i\ln Z[J,\tilde{J}] \,. \label{eq:WFunc}
\end{equation}
For instance, the connected one-point and two-point correlation functions are obtained as\footnote{By slight abuse of notation, we denote the fields' expectation values by the same symbols ($\psi_A$ and $\tilde{\Psi}_A$) as the fields that appear in the path integral \eqref{eq:genFuncSuperfield}. This should not lead to any confusion, however, since from now the path integral will not appear anymore, and thus we always mean the fields' expectation values.}
\begin{align}
    \frac{\delta W[J,\tilde{J}]}{\delta \tilde{J}_A({\Dvec{x}})} = \psi_A({\Dvec{x}}) \,, \quad \frac{\delta W[J,\tilde{J}]}{\delta J_A({\Dvec{x}})} = \tilde{\Psi}_A({\Dvec{x}}) \label{eq:expValsFromW}
\end{align}
and 
\begin{align}
    \frac{\delta^2 W_k[J,\tilde{J}]}{\delta \tilde{J}_A(\Dvec{x}) \delta J_B(\Dvec{y})} &= G_{AB}^R(\Dvec{x},\Dvec{y}) \\
    \frac{\delta^2 W_k[J,\tilde{J}]}{\delta J_A(\Dvec{x}) \delta \tilde{J}_B(\Dvec{y})} &= G_{AB}^A(\Dvec{x},\Dvec{y}) \\
    \frac{\delta^2 W_k[J,\tilde{J}]}{\delta \tilde{J}_A(\Dvec{x}) \delta \tilde{J}_B(\Dvec{y})} &= iF_{AB}(\Dvec{x},\Dvec{y}) \\
    \frac{\delta^2 W_k[J,\tilde{J}]}{\delta J_A(\Dvec{x}) \delta J_B(\Dvec{y})} &= i\widetilde{F}_{AB}(\Dvec{x},\Dvec{y})
\end{align}
where $G^R_{AB}(\Dvec{x},\Dvec{y})$ denotes the retarded propagator, $G^A_{AB}(\Dvec{x},\Dvec{y})$ the advanced propagator, and $iF_{AB}(\Dvec{x},\Dvec{y})$ the statistical function. Causality entails $i\widetilde{F}_{AB}(\Dvec{x},\Dvec{y}) = 0$ for $\tilde{J} = 0$.

The effective action $\Gamma$ is given by the Legendre transform of the Schwinger functional \eqref{eq:WFunc},
\begin{align}
    \Gamma[\psi,\tilde{\Psi}] &=  W[J,\tilde{J}] - \int_{\Dvec{x}} \big( \tilde{J}_A(\Dvec{x}) \psi_A(\Dvec{x}) + J_A(\Dvec{x}) \tilde{\Psi}_A(\Dvec{x}) \big)\, ,  \label{eq:defGam}
\end{align}
where $J=J[\psi,\tilde{\Psi}]$ and $\tilde{J}=\tilde{J}[\psi,\tilde{\Psi}]$ are implicitly determined by inverting \eqref{eq:expValsFromW}. $\Gamma$ is the generator of one-particle irreducible (1PI) correlation functions. 

Importantly, the static flow equation \eqref{eq:staticFlow} is derived by adding the regulator on the level of the LGW functional, 
\begin{equation}
    F \to F + \frac{1}{2}\int \frac{d^dp}{(2\pi)^d} \phi_a(-\vec{p}) R^{\phi}_k(\vec{p})\phi_a(\vec{p}) \label{eq:regInFreeEnergy}
\end{equation}
On the level of the MSR action this amounts to
\begin{align}
    S &\to S + \Delta S_k \quad \label{eq:DeltaSk} \text{with} \\ \nonumber
    &\hspace{1.0cm} \Delta S_k = - \int dt \int \frac{d^dp}{(2\pi)^d} \tilde{\Phi}_a(t,-\vec{p}) R_k^{\phi}(\vec{p}) \phi_a(t,\vec{p}) \,.
\end{align}
Note that the regulator couples to the composite response field $\tilde{\Phi}$ instead of the standard response field $\tilde{\phi}$, which is non-trivial due to the non-linearity of the field transformation \eqref{eq:Jacobian}. As shown in Ref.~\cite{Roth:2024rbi}, this ensures that the relevant symmetries of the MSR action are preserved during the rFRG flow, including $(i)$ the discrete symmetry that expresses thermal equilibrium \cite{Sieberer:2015hba} and $(ii)$ the external temporal gauge symmetry from the Poisson-bracket structure of the ideal part of the equations of motion (which here corresponds to the $O(N)$ generalization of Larmor precession). In particular, $(i)$ entails that the static flow equation \eqref{eq:staticFlow} is unchanged in the presence of dynamics. On the other hand, $(ii)$ ensures that the mode-coupling constant $g$ is protected from renormalization (see also \cite{Canet:2014cta}). 

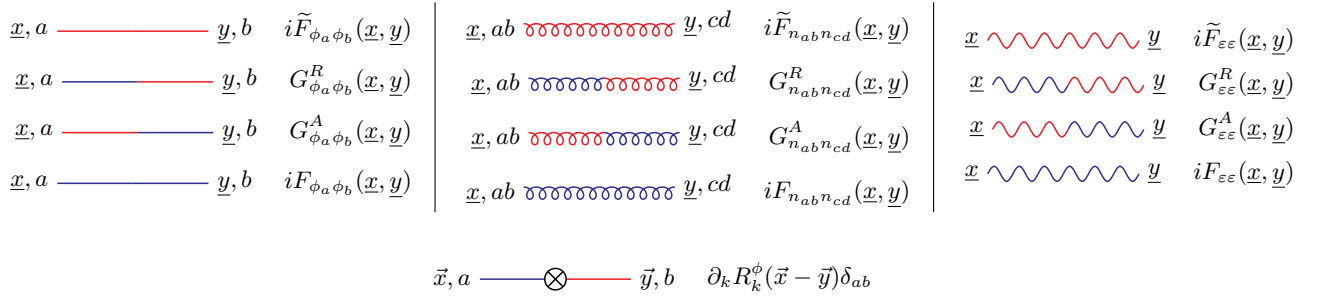
\begin{figure*}
    \centering
    \[
    \begin{array}{c|c|c}
    	\begin{split}
    			\begin{tikzpicture}[baseline=-0.5ex]
    			\draw[pred] (1,0) -- (0,0) node[anchor=east,black] {$\Dvec{x},a$};
    		 	\draw[pred] (2,0) node[anchor=west,black] {$\Dvec{y},b$} -- (1,0);
    		\end{tikzpicture} \quad i\widetilde{F}_{\phi_a \phi_b}(\Dvec{x},\Dvec{y}) \\
    		\begin{tikzpicture}[baseline=-0.5ex,samples=120]
    		      \draw[pblue] (1,0) -- (0,0) node[anchor=east,black] {$\Dvec{x},a$};
    		 	\draw[pred] (2,0) node[anchor=west,black] {$\Dvec{y},b$} -- (1,0);
    		\end{tikzpicture} \quad G^{R}_{\phi_a \phi_b}(\Dvec{x},\Dvec{y})
    		\\
    			\begin{tikzpicture}[baseline=-0.5ex]
    			\draw[pred] (1,0) -- (0,0) node[anchor=east,black] {$\Dvec{x},a$};
    		 	\draw[pblue] (2,0) node[anchor=west,black] {$\Dvec{y},b$} -- (1,0);
    		\end{tikzpicture} \quad G^{A}_{\phi_a \phi_b}(\Dvec{x},\Dvec{y})
    		\\
    			\begin{tikzpicture}[baseline=-0.5ex]
    			\draw[pblue] (1,0) -- (0,0) node[anchor=east,black] {$\Dvec{x},a$};
    		 	\draw[pblue] (2,0) node[anchor=west,black] {$\Dvec{y},b$} -- (1,0);
    		\end{tikzpicture} \quad iF_{\phi_a \phi_b}(\Dvec{x},\Dvec{y})
    	\end{split} \hspace{.2cm}
        &
        \hspace{.2cm}
    	\begin{split}
    		\begin{tikzpicture}[baseline=-0.5ex]
                \draw[pred] node[anchor=east,black] {$\Dvec{x},ab$} plot[domain=0:180, samples=200, smooth, variable=\a] ({0+\a/180 + 0.08*sin(12*\a)},{0.08*cos(12*\a)})  ;
                \draw[pred] plot[domain=0:180, samples=200, smooth, variable=\a] ({1+\a/180 + 0.08*sin(12*\a)},{0.08*cos(12*\a)}) node[anchor=west,black] {$\Dvec{y},cd$} ;
    		\end{tikzpicture} \quad i\widetilde{F}_{n_{ab}n_{cd}}(\Dvec{x},\Dvec{y}) \\
    		\begin{tikzpicture}[baseline=-0.5ex,samples=120]
                \draw[pblue] node[anchor=east,black] {$\Dvec{x},ab$} plot[domain=0:180, samples=200, smooth, variable=\a] ({0+\a/180 + 0.08*sin(12*\a)},{0.08*cos(12*\a)})  ;
                \draw[pred] plot[domain=0:180, samples=200, smooth, variable=\a] ({1+\a/180 + 0.08*sin(12*\a)},{0.08*cos(12*\a)}) node[anchor=west,black] {$\Dvec{y},cd$} ;
    		\end{tikzpicture} \quad G^{R}_{n_{ab}n_{cd}}(\Dvec{x},\Dvec{y})
    		\\
    		\begin{tikzpicture}[baseline=-0.5ex]
                \draw[pred] node[anchor=east,black] {$\Dvec{x},ab$} plot[domain=0:180, samples=200, smooth, variable=\a] ({0+\a/180 + 0.08*sin(12*\a)},{0.08*cos(12*\a)})  ;
                \draw[pblue] plot[domain=0:180, samples=200, smooth, variable=\a] ({1+\a/180 + 0.08*sin(12*\a)},{0.08*cos(12*\a)}) node[anchor=west,black] {$\Dvec{y},cd$} ;
    		\end{tikzpicture} \quad G^{A}_{n_{ab}n_{cd}}(\Dvec{x},\Dvec{y})
    		\\
    		\begin{tikzpicture}[baseline=-0.5ex]
                \draw[pblue] node[anchor=east,black] {$\Dvec{x},ab$} plot[domain=0:180, samples=200, smooth, variable=\a] ({0+\a/180 + 0.08*sin(12*\a)},{0.08*cos(12*\a)})  ;
                \draw[pblue] plot[domain=0:180, samples=200, smooth, variable=\a] ({1+\a/180 + 0.08*sin(12*\a)},{0.08*cos(12*\a)}) node[anchor=west,black] {$\Dvec{y},cd$} ;
    		\end{tikzpicture} \quad iF_{n_{ab}n_{cd}}(\Dvec{x},\Dvec{y})
    	\end{split} \hspace{.2cm}
        &
        \hspace{.2cm}
    	\begin{split}
    		\begin{tikzpicture}[baseline=-0.5ex]
                \draw[pred,smooth,domain=0:1,variable=\x] node[anchor=east,black] {$\Dvec{x}$}  plot (\x,{0.1*sin(6*0.5*2*3.14159*\x r)});
                \draw[pred,smooth,domain=0:1,variable=\x] plot (1+\x,{0.1*sin(6*0.5*2*3.14159*(1+\x) r)}) node[anchor=west,black] {$\Dvec{y}$} ;
    		\end{tikzpicture} \quad i\widetilde{F}_{\varepsilon\varepsilon}(\Dvec{x},\Dvec{y}) \\
    		\begin{tikzpicture}[baseline=-0.5ex,samples=120]
                \draw[pblue,smooth,domain=0:1,variable=\x] node[anchor=east,black] {$\Dvec{x}$}  plot (\x,{0.1*sin(6*0.5*2*3.14159*\x r)});
                \draw[pred,smooth,domain=0:1,variable=\x] plot (1+\x,{0.1*sin(6*0.5*2*3.14159*(1+\x) r)}) node[anchor=west,black] {$\Dvec{y}$} ;
    		\end{tikzpicture} \quad G^{R}_{\varepsilon\varepsilon}(\Dvec{x},\Dvec{y})
    		\\
    		\begin{tikzpicture}[baseline=-0.5ex]
                \draw[pred,smooth,domain=0:1,variable=\x] node[anchor=east,black] {$\Dvec{x}$}  plot (\x,{0.1*sin(6*0.5*2*3.14159*\x r)});
                \draw[pblue,smooth,domain=0:1,variable=\x] plot (1+\x,{0.1*sin(6*0.5*2*3.14159*(1+\x) r)}) node[anchor=west,black] {$\Dvec{y}$} ;
    		\end{tikzpicture} \quad G^{A}_{\varepsilon\varepsilon}(\Dvec{x},\Dvec{y})
    		\\
    		\begin{tikzpicture}[baseline=-0.5ex]
                \draw[pblue,smooth,domain=0:1,variable=\x] node[anchor=east,black] {$\Dvec{x}$}  plot (\x,{0.1*sin(6*0.5*2*3.14159*\x r)});
                \draw[pblue,smooth,domain=0:1,variable=\x] plot (1+\x,{0.1*sin(6*0.5*2*3.14159*(1+\x) r)}) node[anchor=west,black] {$\Dvec{y}$} ;
    		\end{tikzpicture} \quad iF_{\varepsilon\varepsilon}(\Dvec{x},\Dvec{y})
    	\end{split}
        \end{array}
    \]
    \begin{align*}
    \begin{split}
        \begin{tikzpicture}[baseline=-0.5ex,samples=120]
    			\draw[pblue] (1,0) -- (0,0) node[anchor=east,black] {$\vec{x},a$};
    		 	\draw[pred] (2,0) node[anchor=west,black] {$\vec{y},b$} -- (1,0);
                \filldraw[white] (1,0) circle (0.1414213562);
    	        \draw (1,0) circle (0.1414213562);
    		 	\draw[black] (1-0.1,-0.1) -- (1+0.1,+0.1);
    		 	\draw[black] (1-0.1,+0.1) -- (1+0.1,-0.1);
    		\end{tikzpicture} \quad \partial_k R^{\phi}_{k} (\vec{x}-\vec{y}) \delta_{ab} 
    \end{split}
    \end{align*}
    \caption{Diagrammatic representations of the various propagators and the regulator derivative. Color indicates the type of the fields in the MSR formalism, i.e.,~blue denotes the classical fields $\phi$, $n$, and $\varepsilon$, and red the corresponding (composite) response fields $\tilde{\Phi}$, $\tilde{N}$, and $\tilde{\mathcal{E}}$. In addition, a green line denotes the sum over red and blue. Vertices are denoted by black dots.}
    \label{fig:diagrammatics}
\end{figure*}

After adding the regulator, the generating functionals \eqref{eq:genFunc} and \eqref{eq:WFunc} become dependent on the FRG scale, $Z \to Z_k$, $W \to W_k$. One defines the effective \emph{average} action $\Gamma_k$ with an additional subtraction of the regulator term $\Delta S_k$,
\begin{align}
    \Gamma_k[\psi,\tilde{\Psi}] &=  W_k[J,\tilde{J}] \;- \label{eq:defGamk} \\ \nonumber
    &\int_{\Dvec{x}} \big( \tilde{J}_A(\Dvec{x}) \psi_A(\Dvec{x}) + J_A(\Dvec{x}) \tilde{\Psi}_A(\Dvec{x}) \big) -\Delta S_k[\psi,\tilde{\Psi}] \,, 
\end{align}
which is also known as a `modified' Legendre transform.

The scale $k$ dependence of $\Gamma_k$ is determined by the flow equation \cite{Berges:2012ty,Roth:2024rbi}
\begin{align}
    \partial_k \Gamma_k = -\frac{i}{2}\Tr\left\{ \partial_k R_k \tau_1\big(\Gamma_k^{(2)} - R_k\tau_1\big)^{-1} \right\} \label{eq:treeLevelFlowEq}
\end{align}
where $\Gamma_k^{(2)}$ denotes the Hessian of $\Gamma_k$ in the full $(\psi,\tilde{\Psi})$ field space (see below), and $\tau_1$ denotes the first Pauli matrix. The latter appears since the regulator in \eqref{eq:DeltaSk} is off-diagonal in $(\psi,\tilde{\Psi})$ field space,
\begin{align} 
    \Delta S_k = - \frac{1}{2} \int 
    (\psi,\tilde{\Psi})
    \underbrace{\begin{pmatrix}
        0 & R_k \\
        R_k & 0 
    \end{pmatrix}}_{R_k \tau_1}
    \begin{pmatrix}
        \psi\\ \tilde{\Psi}
    \end{pmatrix}
    \,. 
\end{align} 
Denoting functional derivatives by superscripts, 
\begin{align}
    \Gamma^{\psi_A \tilde{\Psi}_B}_k(\Dvec{x},\Dvec{y}) \equiv \frac{\delta^2 \Gamma_k[\psi,\tilde{\Psi}]}{\delta \psi_A(\Dvec{x}) \delta \tilde{\Psi}_B(\Dvec{y})} \quad\text{etc.}\, ,
\end{align}
the components of $\Gamma_k^{(2)}$ read
\begin{align}
    \Gamma_k^{(2)} = \begin{pmatrix}
        \Gamma^{\psi\psi}_k & \Gamma^{\psi\tilde{\Psi}}_k \\
         \Gamma^{\tilde{\Psi}\psi}_k & \Gamma^{\tilde{\Psi}\tilde{\Psi}}_k
    \end{pmatrix} \,.
\end{align}
As such, the flow equation \eqref{eq:treeLevelFlowEq} involves the scale $k$ and field dependent propagator $G_k= -(\Gamma_k^{(2)} - R_k\tau_{1})^{-1}$, whose components
\begin{equation}
    G_k = \begin{pmatrix}
        iF_k & G_k^R \\
        G_k^A & i\widetilde{F}_k
    \end{pmatrix}
\end{equation}
are given by
\begin{align*}
    G_k^R &= -\left\{ (\Gamma^{\tilde{\Psi}\psi}_k-R_k) - \Gamma^{\tilde{\Psi}\tilde{\Psi}}_k  (\Gamma^{\psi\tilde{\Psi}}_k-R_k)^{-1} \Gamma^{\psi\psi}_k \right\}^{-1} \!\! \,, \\
    G_k^A &= -\left\{ (\Gamma^{\psi\tilde{\Psi}}_k-R_k) - \Gamma^{\psi\psi}_k  (\Gamma^{\tilde{\Psi}\psi}_k-R_k)^{-1}  \Gamma^{\tilde{\Psi}\tilde{\Psi}}_k \right\}^{-1} \!\! \,, \\
    iF_k &= -\left\{ \Gamma^{\psi\psi}_k - (\Gamma^{\psi\tilde{\Psi}}_k-R_k)   (\Gamma^{\tilde{\Psi}\tilde{\Psi}}_k)^{-1}  (\Gamma^{\tilde{\Psi}\psi}_k-R_k) \right\}^{-1} \!\! \,, \\
    i\widetilde{F}_k &= -\left\{ \Gamma^{\tilde{\Psi}\tilde{\Psi}}_k - (\Gamma^{\tilde{\Psi}\psi}_k-R_k)   (\Gamma^{\psi\psi}_k)^{-1}  (\Gamma^{\psi\tilde{\Psi}}_k-R_k) \right\}^{-1} \!\! \, .
\end{align*} 
These become particularly simple for vanishing response fields, $\tilde{\Psi}=0$, since in this case causality requires $\Gamma^{\psi\psi}=0$, and one is left with 
\begin{align*}
    G_k^R &= -(\Gamma^{\tilde{\Psi}\psi}_k-R_k)^{-1} \,, \hspace{1.0cm} iF_k = G_k^R \Gamma^{\tilde{\Psi}\tilde{\Psi}}_k G_k^A \,, \\
    G_k^A &= - (\Gamma^{\psi\tilde{\Psi}}_k-R_k)^{-1} \,, \hspace{1.0cm}
    i\widetilde{F}_k = 0 \,.
\end{align*} 

Our diagrammatic representations for the various propagators and regulator derivatives are shown in Fig.~\ref{fig:diagrammatics}. With this notation, the flow equation \eqref{eq:treeLevelFlowEq} is compactly expressed as a one-loop diagram,
\begin{equation}
    \partial_k \Gamma_k = \label{eq:treeLevelFlowDiag}
    \frac{i}{2} 
    \;
        \begin{tikzpicture}[baseline=-0.5ex]
            \draw[pgreen] plot[domain=0:360, samples=200, smooth, variable=\a] ({0.5*cos(\a)},{0.5*sin(\a)});
            \filldraw[white] (0,0.5) circle (0.1414213562);
            \draw (0,0.5) circle (0.1414213562);
            \draw[black] (-0.1,0.5-0.1) -- (+0.1,0.5+0.1);
            \draw[black] (-0.1,0.5+0.1) -- (+0.1,0.5-0.1);
        \end{tikzpicture}\;
\end{equation}
The regulators for the energy-like density and the $O(N)$ charge densities are set to zero, $R_k^{\varepsilon} = R_k^{n} = 0$. Otherwise, their fluctuations would appear as additional loops.

We truncate the effective average action by promoting the kinetic coefficients $\Gamma^{\phi}$, $\gamma$ and $\mu$ to depend on the FRG scale $k$,
\begin{widetext}
\begin{align}
    \Gamma_k = \int_{\Dvec{x}} \bigg\{ &-\tilde{\phi}_a\left(\frac{\partial \phi_a}{\partial t} + \Gamma_k^{\phi} \frac{\delta F_k}{\delta \phi_a} - \frac{g}{2} \{ \phi_a, n_{bc} \} \frac{\delta F_k}{\delta n_{bc}} \right) \nonumber \\
    &-\frac{1}{2}\tilde{n}_{ab} \left( \frac{\partial n_{ab}}{\partial t} - \gamma_k \vec{\nabla}^2 \frac{\delta F_k}{\delta n_{ab}} - g\{n_{ab},\phi_c\} \frac{\delta F_k}{\delta \phi_c} - \frac{g}{2} \{ n_{ab},n_{cd} \}\frac{\delta F_k}{\delta n_{cd}} \right) \nonumber \\
    &-\tilde{\varepsilon} \left( \frac{\partial \varepsilon}{\partial t} - \mu_k \vec{\nabla}^2 \frac{\delta F_k}{\delta \varepsilon} \right)  + iT \Gamma^{\phi}_k \tilde{\phi}_a\tilde{\phi}_a - \frac{i}{2} T  \tilde{n}_{ab} \gamma_k \vec{\nabla}^2 \tilde{n}_{ab} - i T  \tilde{\varepsilon} \mu_k \vec{\nabla}^2 \tilde{\varepsilon} \bigg\} \, . \label{eq:EffAvgAction}
\end{align}
\end{widetext}
Note that $\Gamma_k$ depends on the \emph{composite} response fields $\tilde{\Phi}_a$, $\tilde{N}_{ab}$, and $\tilde{\mathcal{E}}$.  As such, the standard response fields $\tilde{\phi}_a$, $\tilde{n}_{ab}$, and $\tilde{\varepsilon}$ appearing in \eqref{eq:EffAvgAction} are implicitly defined as the inverse of the generalization of \eqref{eq:ResponseFields} to finite FRG scales~$k$, i.e.~by,
\begin{align}
    \tilde{\Phi}_a &= \Gamma^{\phi}_k \tilde{\phi}_a + \frac{g}{2}\{\phi_a,n_{bc}\} \tilde{n}_{ab}   \label{eq:ResponseFieldsFRGFlow} \\ \nonumber
    \tilde{N}_{ab} &= -\gamma_k \vec{\nabla}^2 \tilde{n}_{ab} + g\{n_{ab},\phi_c\} \tilde{\phi}_c  + \frac{g}{2} \{n_{ab},n_{cd}\} \tilde{n}_{cd}  \\ \nonumber
    \tilde{\mathcal{E}} &= -\mu_k\vec{\nabla}^2 \tilde{\varepsilon} 
\end{align} 

By virtue of the fluctuation-dissipation relation, the kinetic coefficients appear in the variances of the noises \eqref{eq:noiseVariances}, and can thus be extracted from the effective average action $\Gamma_k$ via second functional derivatives with respect to the (composite) response fields (no sum over $a$ and $ab$),
\begin{align}
    \frac{1}{\Gamma^{\phi}_k} &= \frac{1}{2iT} \lim_{\vec{p},\omega\to 0}  \frac{\delta^2 \Gamma_k}{\delta \tilde{\Phi}_a(-\omega,-\vec{p})\delta \tilde{\Phi}_a(\omega,\vec{p})} \bigg\rvert_{0} \, , \nonumber \\
    \frac{1}{\gamma_k} &= \frac{1}{2iT} \lim_{\vec{p}\to 0} \vec{p}^{\,2} \lim_{\omega\to 0} \frac{\delta^2 \Gamma_k}{\delta \tilde{N}_{ab}(-\omega,-\vec{p})\delta \tilde{N}_{ab}(\omega,\vec{p})} \bigg\rvert_{0} \, , \nonumber \\
    \frac{1}{\mu_k} &= \frac{1}{2iT} \lim_{\vec{p}\to 0} \vec{p}^{\,2} \lim_{\omega\to 0} \frac{\delta^2 \Gamma_k}{\delta \tilde{\mathcal{E}}(-\omega,-\vec{p})\delta \tilde{\mathcal{E}}(\omega,\vec{p})} \bigg\rvert_{0} \, , \label{eq:defSigmak}
\end{align}
where the right-hand side is evaluated at vanishing field expectation values, $\psi=\tilde{\Psi}=0$. 

\newcommand{\sA}{\alpha} 
\newcommand{\sB}{\beta}

Functional derivatives of the flow equation \eqref{eq:treeLevelFlowDiag} can be efficiently performed diagrammatically \cite{Huelsmann:2020xcy}. For example, the rather lengthy expression
\begin{align}
    \frac{\delta}{\delta \tilde{\Phi}_a(\Dvec{z})} &G^{R}_{\varepsilon\varepsilon}(\Dvec{x},\Dvec{y}) = \\ \int_{\Dvec{v}\Dvec{w}} \Big[ &G^{R}_{\varepsilon \psi_{\sA}}(\Dvec{x},\Dvec{v}) \Gamma^{\tilde{\Psi}_{\sA} \tilde{\Phi}_a \psi_{\sB}}(\Dvec{v},\Dvec{z},\Dvec{w}) G^{R}_{\psi_{\sB} \varepsilon}(\Dvec{w},\Dvec{y}) \nonumber \\ +\; &G^{R}_{\varepsilon \psi_{\sA}}(\Dvec{x},\Dvec{v}) \Gamma^{\tilde{\Psi}_{\sA} \tilde{\Phi}_a \tilde{\Psi}_{\sB}}(\Dvec{v},\Dvec{z},\Dvec{w}) i\widetilde{F}_{\psi_{\sB} \varepsilon}(\Dvec{w},\Dvec{y}) \nonumber \\ 
    +\; & iF_{\varepsilon \psi_{\sA}}(\Dvec{x},\Dvec{v}) \Gamma^{\psi_{\sA} \tilde{\Phi}_a \tilde{\Psi}_{\sB}}(\Dvec{v},\Dvec{z},\Dvec{w}) i\widetilde{F}_{\psi_{\sB} \varepsilon}(\Dvec{w},\Dvec{y})  \nonumber \\ +\; &iF_{\varepsilon \psi_{\sA}}(\Dvec{x},\Dvec{v}) \Gamma^{\psi_{\sA} \tilde{\Phi}_a \psi_{\sB}}(\Dvec{v},\Dvec{z},\Dvec{w}) G^R_{\psi_{\sB} \varepsilon}(\Dvec{w},\Dvec{y}) \Big] \nonumber
\end{align}
would be compactly denoted as
\begin{align*}
    \frac{\delta}{\delta \tilde{\Phi}_a(\Dvec{z})} G^{R}_{\varepsilon \varepsilon}(\Dvec{x},\Dvec{y}) =
    \begin{tikzpicture}[baseline=-0.5ex,samples=120]
        \draw[pblue,smooth,domain=0:0.8,variable=\x] node[anchor=east,black] {$\Dvec{x}$}  plot (\x,{0.1*sin(2*2*3.14159*\x/0.8 r)});
        \draw[pred,smooth,domain=0:0.8,variable=\x] plot (2+\x,{0.1*sin(2*2*3.14159*(1+\x/0.8) r)}) node[anchor=west,black] {$\Dvec{y}$} ;
        \draw[pgreen,line width=2mm] (0.8,0) -- (2,0);
        \draw[pred] (1.4,-0.1) -- (1.4,-0.7) node[anchor=west,black] {$\Dvec{z},a$};
    \end{tikzpicture} 
\end{align*}
where the intermediate thick green line denotes the sum over all possible field ($\phi_a,n_{ab},\varepsilon$) and color (red, blue) permutations. This shorthand notation allows one to evaluate functional derivatives of the flow equation \eqref{eq:treeLevelFlowDiag} with a drastically reduced number of diagrams.

\begin{figure*}
    \centering
\begin{align*}
    \partial_k \Gamma_k^{\tilde{\Phi}_a\tilde{\Phi}_b}(\Dvec{x},\Dvec{y}) &=
    i\;\bigg\{ 
        \begin{tikzpicture}[baseline=-0.5ex]
            \draw[pblue] plot[domain=0:45, samples=200, smooth, variable=\a] ({0.5*cos(\a)},{0.5*sin(\a)});
            \draw[pgreen] plot[domain=45:135, samples=200, smooth, variable=\a] ({0.5*cos(\a)},{0.5*sin(\a)});
            \draw[pblue] plot[domain=135:180, samples=200, smooth, variable=\a] ({0.5*cos(\a)},{0.5*sin(\a)});
            \draw[pblue] plot[domain=180:360, samples=200, smooth, variable=\a] ({(0.5-0.08)*cos(\a)+0.08*cos(15*\a)},{(0.5-0.08)*sin(\a)+0.08*sin(15*\a)-0.05});
    	 	\draw[pred] (-0.6,0) -- (-0.6-0.5,0);
    	 	\draw[pred] (0.5,0) -- (0.6+0.5,0);
            \node[] at (-0.9,-0.25) {\footnotesize $\Dvec{x},a$};
            \node[] at (+0.9,-0.25) {\footnotesize $\Dvec{y},b$};
		 	\fill[black] (-0.5,0) circle (0.1);
		 	\fill[black] (+0.5,0) circle (0.1);
            \filldraw[white] (0,0.5) circle (0.1414213562);
            \draw (0,0.5) circle (0.1414213562);
            \draw[black] (-0.1,0.5-0.1) -- (+0.1,0.5+0.1);
            \draw[black] (-0.1,0.5+0.1) -- (+0.1,0.5-0.1);
        \end{tikzpicture}  +\frac{1}{2} 
        \begin{tikzpicture}[baseline=-0.5ex]
            \draw[pgreen] plot[domain=0:180, samples=200, smooth, variable=\a] ({0.5*cos(\a)},{0.5*sin(\a)});
            \draw[pblue] plot[domain=180:360, samples=200, smooth, variable=\a] ({0.5*cos(\a)},{0.5*sin(\a)});
            \draw[pred] (-0.1,-0.5) -- (-0.6,-0.5) ;
            \draw[pred] (0.1,-0.5) -- (+0.6,-0.5);
            \node[] at (-0.4,-0.75) {\footnotesize $\Dvec{x},a$};
            \node[] at (+0.4,-0.75) {\footnotesize $\Dvec{y},b$};
            \fill[black] (0,-0.5) circle (0.1);
            \filldraw[white] (0,0.5) circle (0.1414213562);
            \draw (0,0.5) circle (0.1414213562);
            \draw[black] (-0.1,0.5-0.1) -- (+0.1,0.5+0.1);
            \draw[black] (-0.1,0.5+0.1) -- (+0.1,0.5-0.1);
        \end{tikzpicture} + 
        \begin{tikzpicture}[baseline=-0.5ex]
            \draw[pblue] plot[domain=0:45, samples=200, smooth, variable=\a] ({0.5*cos(\a)},{0.5*sin(\a)});
            \draw[pgreen] plot[domain=45:135, samples=200, smooth, variable=\a] ({0.5*cos(\a)},{0.5*sin(\a)});
            \draw[pblue] plot[domain=135:180, samples=200, smooth, variable=\a] ({0.5*cos(\a)},{0.5*sin(\a)});
            \draw[pblue] plot[domain=180:360, samples=200, smooth, variable=\a] ({0.5*cos(\a)*(1.0-0.1*cos(10*\a))},{0.5*sin(\a)*(1.0-0.1*cos(10*\a))});
    	 	\draw[pred] (-0.5,0) -- (-0.5-0.5,0);
    	 	\draw[pred] (0.5,0) -- (0.5+0.5,0);
            \node[] at (-0.9,-0.25) {\footnotesize $\Dvec{x},a$};
            \node[] at (+0.9,-0.25) {\footnotesize $\Dvec{y},b$};
		 	\fill[black] (-0.5,0) circle (0.05);
		 	\fill[black] (+0.5,0) circle (0.05);
            \filldraw[white] (0,0.5) circle (0.1414213562);
            \draw (0,0.5) circle (0.1414213562);
            \draw[black] (-0.1,0.5-0.1) -- (+0.1,0.5+0.1);
            \draw[black] (-0.1,0.5+0.1) -- (+0.1,0.5-0.1);
        \end{tikzpicture} 
    \bigg\} \\
    \partial_k \Gamma_k^{\tilde{N}_{ab}\tilde{N}_{cd}}(\Dvec{x},\Dvec{y}) &=
    i\;\bigg\{ 
        \begin{tikzpicture}[baseline=-0.5ex]
            \draw[pblue] plot[domain=0:45, samples=200, smooth, variable=\a] ({0.5*cos(\a)},{0.5*sin(\a)});
            \draw[pgreen] plot[domain=45:135, samples=200, smooth, variable=\a] ({0.5*cos(\a)},{0.5*sin(\a)});
            \draw[pblue] plot[domain=135:180, samples=200, smooth, variable=\a] ({0.5*cos(\a)},{0.5*sin(\a)});
            \draw[pblue] plot[domain=180:360, samples=200, smooth, variable=\a] ({0.5*cos(\a)},{0.5*sin(\a)});
            \draw[pred] plot[domain=0:90, samples=200, smooth, variable=\a] ({-1.1+\a/180 + 0.08*sin(12*\a)},{0.08*cos(12*\a) - 0.08});
            \draw[pred] plot[domain=0:90, samples=200, smooth, variable=\a] ({+0.6+\a/180 + 0.08*sin(12*\a)},{0.08*cos(12*\a) - 0.08});
            \node[] at (-0.9,-0.35) {\footnotesize $\Dvec{x},ab$};
            \node[] at (+0.9,-0.35) {\footnotesize $\Dvec{y},cd$};
		 	\fill[black] (-0.5,0) circle (0.1);
		 	\fill[black] (+0.5,0) circle (0.1);
            \filldraw[white] (0,0.5) circle (0.1414213562);
            \draw (0,0.5) circle (0.1414213562);
            \draw[black] (-0.1,0.5-0.1) -- (+0.1,0.5+0.1);
            \draw[black] (-0.1,0.5+0.1) -- (+0.1,0.5-0.1);
        \end{tikzpicture} + \frac{1}{2}
        \begin{tikzpicture}[baseline=-0.5ex]
            \draw[pgreen] plot[domain=0:180, samples=200, smooth, variable=\a] ({0.5*cos(\a)},{0.5*sin(\a)});
            \draw[pblue] plot[domain=180:360, samples=200, smooth, variable=\a] ({0.5*cos(\a)},{0.5*sin(\a)});
            \draw[pred] plot[domain=0:90, samples=200, smooth, variable=\a] ({-0.6+\a/180 + 0.08*sin(12*\a)},{0.08*cos(12*\a) - 0.08 - 0.5});
            \draw[pred] plot[domain=0:90, samples=200, smooth, variable=\a] ({0.1+\a/180 + 0.08*sin(12*\a)},{0.08*cos(12*\a) - 0.08 - 0.5});
            \node[] at (-0.4,-0.85) {\footnotesize $\Dvec{x},ab$};
            \node[] at (+0.4,-0.85) {\footnotesize $\Dvec{y},cd$};
            \fill[black] (0,-0.5) circle (0.1);
            \filldraw[white] (0,0.5) circle (0.1414213562);
            \draw (0,0.5) circle (0.1414213562);
            \draw[black] (-0.1,0.5-0.1) -- (+0.1,0.5+0.1);
            \draw[black] (-0.1,0.5+0.1) -- (+0.1,0.5-0.1);
        \end{tikzpicture}
    \bigg\} \\
    \partial_k \Gamma_k^{\tilde{\mathcal{E}}\tilde{\mathcal{E}}}(\Dvec{x},\Dvec{y}) &=
    i\;\bigg\{ 
        \begin{tikzpicture}[baseline=-0.5ex]
            \draw[pblue] plot[domain=0:45, samples=200, smooth, variable=\a] ({0.5*cos(\a)},{0.5*sin(\a)});
            \draw[pgreen] plot[domain=45:135, samples=200, smooth, variable=\a] ({0.5*cos(\a)},{0.5*sin(\a)});
            \draw[pblue] plot[domain=135:180, samples=200, smooth, variable=\a] ({0.5*cos(\a)},{0.5*sin(\a)});
            \draw[pblue] plot[domain=180:360, samples=200, smooth, variable=\a] ({0.5*cos(\a)},{0.5*sin(\a)});
            \draw[pred,smooth,domain=0:0.5,variable=\x] plot (-\x-0.5,{0.1*sin(6*0.5*2*3.14159*\x r)});
            \draw[pred,smooth,domain=0:0.5,variable=\x] plot (\x+0.5,{0.1*sin(6*0.5*2*3.14159*\x r)});
            \node[] at (-0.9,-0.25) {\footnotesize $\Dvec{x}$};
            \node[] at (+0.9,-0.25) {\footnotesize $\Dvec{y}$};
		 	\fill[black] (-0.5,0) circle (0.05);
		 	\fill[black] (+0.5,0) circle (0.05);
            \filldraw[white] (0,0.5) circle (0.1414213562);
            \draw (0,0.5) circle (0.1414213562);
            \draw[black] (-0.1,0.5-0.1) -- (+0.1,0.5+0.1);
            \draw[black] (-0.1,0.5+0.1) -- (+0.1,0.5-0.1);
        \end{tikzpicture} 
        \bigg\}
\end{align*}
    \caption{Flow of statistical 2-point functions at vanishing field expectation values $\psi=\tilde{\Psi}=0$. The tree-level vertices for the interactions of $\phi_{a}$ and $n_{ab}$ are non-local due to the field transformation \eqref{eq:ResponseFieldsFRGFlow}, which we emphasize by a bold black dot. The corresponding Feynman rules are given in Appendix~\ref{sec:1PIVertFnc}. In particular, the $\phi\phi\tilde{\Phi}\tilde{\Phi}$ vertex appearing in the tadpole diagram in the first line is proportional to $g^2$ and thus also represents an interaction with the charge densities $n_{ab}$.}
    \label{fig:flowOf2PtFnc}
\end{figure*}

For the flow equations of the statistical two-point functions $\partial_k \Gamma_k^{\tilde{\Phi}\tilde{\Phi}}$, $\partial_k \Gamma_k^{\tilde{N}\tilde{N}}$ and $\partial_k \Gamma_k^{\tilde{\mathcal{E}}\tilde{\mathcal{E}}}$ at vanishing field expectation values,  this procedure leads to the diagrams depicted in Fig.~\ref{fig:flowOf2PtFnc}. To evaluate these diagrams via Feynman rules, we need expressions for the vertex functions and the propagators within our truncation. First, the propagators are given by (at the expansion point $\psi=\tilde{\Psi}=0$),
\begin{align}
	G^{R}_{\phi_a\phi_b,k}(\omega,\vec{p})
    &= -\frac{\Gamma_k^{\phi} \delta_{ab}}{+i\omega-\Gamma_k^{\phi} (m_k^2 + \vec{p}^2 + R_k^{\phi}(\vec{p}) )} \,, \label{eq:GRPhi} \\
	G^{A}_{\phi_a\phi_b,k}(\omega,\vec{p})
    &= -\frac{\Gamma_k^{\phi} \delta_{ab}}{-i\omega-\Gamma_k^{\phi} (m_k^2 + \vec{p}^2 + R_k^{\phi}(\vec{p}) )} \,, \label{eq:GAPhi} \\
	iF_{\phi_a\phi_b,k}(\omega,\vec{p}) &= \frac{2i\Gamma_k^{\phi} T \delta_{ab}}{\omega^2 + (\Gamma_k^{\phi} )^2\,(m_k^2 + \vec{p}^2 + R_k^{\phi}(\vec{p}) )^2} \,, \label{eq:iFPhi}\\
	i\widetilde{F}_{\phi_a\phi_b,k}(\omega,\vec{p}) &= 0 \,, \label{eq:iFTildePhi}
\end{align}
\begin{align}
	G^{R}_{n_{ab}n_{cd},k}(\omega,\vec{p})
    &= -\frac{\gamma_k \vec{p}^2 (\delta_{ac}\delta_{bd}-\delta_{ad}\delta_{bc})}{+i\omega-\gamma_k \vec{p}^2/\chi_n} \,, \label{eq:GRn} \\
	G^{A}_{n_{ab}n_{cd},k}(\omega,\vec{p})
    &= -\frac{\gamma_k \vec{p}^2 (\delta_{ac}\delta_{bd}-\delta_{ad}\delta_{bc})}{-i\omega-\gamma_k \vec{p}^2/\chi_n} \,, \label{eq:GAn} \\
	iF_{n_{ab}n_{cd},k}(\omega,\vec{p}) &= \frac{2i\gamma_k \vec{p}^2 T (\delta_{ac}\delta_{bd}-\delta_{ad}\delta_{bc})}{\omega^2 + (\gamma_k \vec{p}^2/\chi_n)^2} \,, \label{eq:iFn}\\
	i\widetilde{F}_{n_{ab}n_{cd},k}(\omega,\vec{p}) &= 0 \,, \label{eq:iFTilden}
\end{align}
and
\begin{align}
	G^{R}_{\varepsilon\varepsilon,k}(\omega,\vec{p})
    &= -\frac{\mu_k \vec{p}^2}{+i\omega-\mu_k \vec{p}^2/\chi_{\varepsilon,k}} \,, \label{eq:GRe} \\
	G^{A}_{\varepsilon\varepsilon,k}(\omega,\vec{p})
    &= -\frac{\mu_k \vec{p}^2}{-i\omega-\mu_k \vec{p}^2/\chi_{\varepsilon,k}} \,, \label{eq:GAe} \\
	iF_{\varepsilon\varepsilon,k}(\omega,\vec{p}) &= \frac{2i\mu_k \vec{p}^2 T}{\omega^2 + (\mu_k \vec{p}^2/\chi_{\varepsilon,k})^2} \,, \label{eq:iFe} \\
	i\widetilde{F}_{\varepsilon\varepsilon,k}(\omega,\vec{p}) &= 0 \,. \label{eq:iFTildee}
\end{align}
To obtain the Feynman rules for the vertex functions, we need to invert the Jacobian matrix \eqref{eq:Jacobian} for inhomogeneous field expectation values, which is generally not possible in closed form. Importantly, the two blocks in the Jacobian matrix \eqref{eq:Jacobian} can be inverted separately.
The inversion of the upper left block matrix was already performed in Ref.~\cite{Roth:2024rbi} in a Neumann series up to second order in the classical fields $\phi_a$ and $n_{ab}$ which is sufficient for obtaining the 3-point and 4-point vertices. Hence, the 1PI vertex functions containing only the order parameter and the $O(N)$ charge densities are the same as the ones listed in Appendix~C.4 of Ref.~\cite{Roth:2024rbi}. For the convenience of the reader, they are here provided in Appendix~\ref{sec:1PIVertFnc}.
On the other hand, the vertices that contain $\phi$ and $\varepsilon$ are given by
\begin{align}
    &\Gamma_k^{\tilde{\Phi}_a \phi_b \varepsilon}(\Dvec{p},\Dvec{q},\Dvec{r}) = \Gamma_k^{\tilde{\mathcal{E}} \phi_a \phi_b}(\Dvec{p},\Dvec{q},\Dvec{r}) = \label{eq:Epp-vertex} 
    \\ \nonumber
    &\hspace{1.0cm} -\frac{\iota_k}{\sqrt{N}} ~ \delta_{ab} ~ (2\pi)^{d+1} \delta(\Dvec{p}+\Dvec{q}+\Dvec{r}) \,.
\end{align}
where the $\delta$-distribution ensures overall frequency and wavenumber conservation. On the level of the flow equation for the statistical 2-point function of the order parameter $\Gamma_k^{\tilde{\Phi}_a\tilde{\Phi}_b}$, depicted in Fig.~\ref{fig:flowOf2PtFnc}, the diagrams are cleanly separated into contributions from the $O(N)$ charge densities and the energy-like density, respectively, allowing the flow of its kinetic coefficient to be written as 
\begin{equation}
    \partial_k \Gamma_k^{\phi} = \partial_k  \Gamma_k^{\phi}\big\rvert_{n} +\partial_k \Gamma_k^{\phi} \big\rvert_{\varepsilon} \label{eq:dGammaPhidk}
\end{equation}
The calculation of the two diagrams containing the $O(N)$ charge densities was already performed in Ref.~\cite{Roth:2024hcu}, 
with the result given in the corresponding Supplemental Material,
\begin{align}
    \partial_k  \Gamma_k^{\phi}\big\rvert_{n} &= \frac{g^2 \, (N-1)  \,d \,\Omega _d k^{d-1} T\,\Gamma _k^{\phi }}{(2\pi)^{d} \, \left(k^2+m_k^2\right) \, \gamma_k} \times \label{Eq:dGammadkN} \\ \nonumber
    &\hspace{0.5cm} \Bigg\{\frac{ 1 
   }{k^2\gamma _k/\chi_n+  \Gamma _k^{\phi
   } ( k^2+ m_k^2 ) } \; - \\ \nonumber
   &\hspace{1.0cm}\frac{2+(d-4) \; {}_2F_1\left(1,\frac{d-2}{2};\frac{d}{2};-\frac{k^2 \gamma _k/\chi_n}{\Gamma _k^{\phi
   } (k^2+m_k^2 )}\right)}{(d-2)\,\Gamma _k^{\phi }\left(k^2+m_k^2\right)}\Bigg\}  \, ,
\end{align}
involving the hypergeometric function ${}_2F_1$.
As such, we only need to perform the calculation of the third diagram of the corresponding flow equation in Fig.~\ref{fig:flowOf2PtFnc}, here. Following the Feynman rules, with Eq.~\eqref{eq:Epp-vertex}, this diagram evaluates to 
\begin{align}
    I &= -\frac{i\iota_k^2}{N}\int \frac{d^dp}{(2\pi)^d}  \int \frac{d\omega}{2\pi} B^F_{\phi_a\phi_b}(\omega,\vec{p}) iF_{\varepsilon\varepsilon}(\omega,\vec{p}) \, ,
\end{align}
where
\begin{align}
    B^F_{\phi_a\phi_b}(\omega,\vec{p}) &= G^R_{\phi_a \phi_c}(\omega,\vec{p}) \partial_k R_k(\vec{p}) iF_{\phi_c \phi_b}(\omega,\vec{p}) \\ \nonumber 
    &+ iF_{\phi_a \phi_c}(\omega,\vec{p}) \partial_k R_k(\vec{p}) G^A_{\phi_c \phi_b}(\omega,\vec{p}) \,.
\end{align}
We perform the frequency integral with the residue theorem, and the momentum integral analytically (which is possible due to the convenient form of the optimized regulator \eqref{eq:RPhi}), and find
\begin{widetext}
\begin{align}
    \partial_k \Gamma_k^{\phi} \big\rvert_{\varepsilon} &= \frac{\iota_k^2 \chi_{\varepsilon,k}\,d\,\Omega_d k^{d+1} T (\Gamma_k^{\phi})^2}{(2\pi)^d \, (k^2+m_k^2)^2\,N} \Bigg\{ \frac{1}{\Gamma_k^{\phi}(k^2+m_k^2)} + 
   \frac{1}{\mu_k k^2/\chi_{\varepsilon,k}+\Gamma_k^{\phi}(k^2+m_k^2)} \label{Eq:dGammadkE} \\ \nonumber 
   &\hspace{3.5cm} - \frac{d\,\mu_k k^2/\chi_{\varepsilon,k}}{(2+d) (\Gamma_k^{\phi})^2(k^2+m_k^2)^2} \; {}_2F_1\left( 1,1+\frac{d}{2}; 2+\frac{d}{2};-\frac{\mu_k k^2/\chi_{\varepsilon,k}}{\Gamma_k^{\phi}(k^2+m_k^2)} \right)  \\ \nonumber
   &\hspace{3.5cm}  - \frac{2(d-2) }{d\,\Gamma_k^{\phi}(k^2+m_k^2)} \; {}_2F_1\left( 1,\frac{d}{2};1+\frac{d}{2}; -\frac{\mu_k k^2/\chi_{\varepsilon,k}}{\Gamma_k^{\phi}(k^2+m_k^2)} \right)
   \Bigg\} \, .
\end{align}
\end{widetext}

The $O(N)$ charge densities do not couple to the energy-like density directly, and thus the flow equation for $\gamma_k$ is the same as in Ref.~\cite{Roth:2024hcu},
\begin{align}
   \partial_k \gamma_k &= -\frac{2 g^2 \Omega _d k^{d+1} T}{(2\pi)^d \, \Gamma _k^{\phi } \,  \left(k^2+m_k^2\right)^3}\,. \label{Eq:dgammadk}
\end{align}

Finally, one can straightforwardly verify that the diagram in the last line of Fig.~\ref{fig:flowOf2PtFnc} is finite for $\vec{p}\to 0$. With the additional factor of $\vec{p}^2$ in the projection \eqref{eq:defSigmak}, the flow of $\mu_k$ vanishes,
\begin{eqnarray}
\label{Eq:sigma_kFlow}
k\partial_k \mu_k = 0\,\,, 
\end{eqnarray}
analogous to Model~C \cite{Halperin:1974zz}. 

\section{Results}
\label{sec:results}

\subsection{Statics}
\label{sec:resultsStatics}

To discuss the fixed points of the static FRG flow, we introduce the dimensionless couplings $\bar{m}^2 = k^{-2} m_k^2$, $\bar{\lambda}' = k^{d-4} T\lambda_k'$, $\bar{\kappa} = k^{2d-6} T^2 \kappa_k$, $\bar{v} = k^{d-4} T \iota_k^2 \chi_{\varepsilon,k}$, where the powers of $k$ correspond to the canonical scaling dimension of the couplings. The powers of $T$ are needed for $F/T$ to be dimensionless.  
In particular, $\bar{v}$ describes the effective coupling of the order parameter to the energy-like density. 
In dimensionless form, the flow equations in (\ref{eq:m2LambdaKappaFlow}) become
\begin{align}
    k\partial_k \bar{m}^2 &= -2\bar{m}^2 - 
    \frac{K_d(N+2) \bar{\lambda}'}{3N (1+\bar{m}^2)^2}  \label{eq:m2LambdaKappaBarFlow}\, , \\ \nonumber
    k\partial_k \bar{\lambda}' &= -(4-d) \bar{\lambda}' + 
    \frac{2K_d(N+8) \bar{\lambda}'^2}{3N (1+\bar{m}^2)^3} - 
    \frac{K_d(N+4)\bar{\kappa}}{5N (1+\bar{m}^2)^2} \, , \\ \nonumber
    k\partial_k \bar{\kappa} &= -2(3-d)\bar{\kappa} \\ \nonumber
    &\hspace{1.1cm} + 
    \frac{2K_d(N+14) \bar{\lambda}'\bar{\kappa}}{N(1+\bar{m}^2)^3} - 
    \frac{10K_d(N+26) \bar{\lambda}'^3}{3N(1+\bar{m}^2)^4} \, . 
\end{align}
On the other hand, the flow equations (\ref{eq:iotaFlow}) and (\ref{eq:chiFlow}) for $\iota_k$ and $\chi_{\varepsilon,k}$ reduce to a single equation for $\bar{v}$, 
\begin{equation}
    k\partial_k \bar{v} = -(4-d)\bar{v} + 
    \frac{2K_d\bar{v}(2(N+2)\bar{\lambda}' + 3N\bar{v})}{3N(1+\bar{m}^2)^3} \, . \label{eq:vBarFlow}
\end{equation}
In particular, using the shifted quartic coupling, in dimensionless units here given by $\bar{\lambda'}=\bar{\lambda}-3\bar{v}$, conveniently removes the dependence on $\bar{v}$ from Eqs.~\eqref{eq:m2LambdaKappaBarFlow}, such that these can be solved independent of \eqref{eq:vBarFlow}. This is because one could have integrated out the quadratic energy density in \eqref{eq:freeEnergy}, which would correspond to a shift in the quartic coupling, $\lambda \to \lambda' = \lambda - 3\iota^2 \chi_{\varepsilon}$.  

The upper critical dimension of the $(\phi^2)^3$ coupling in LGW functional \eqref{eq:freeEnergy} is $d=3$, with marginally irrelevant coupling $\kappa$. In this case, the tricritical point corresponds to a Gaussian fixed point of the FRG flow which, in general spatial dimension $d$, is located at\footnote{These fixed-point values (particularly $\bar{v}^*$) stay the same when the full effective potential $U_k(\rho,\varepsilon)$ is taken into account, as shown in Appendix~\ref{sec:exactSolEffPotGaussFP}.}  
\begin{equation}
    \bar{m}^{2*} = \bar{\lambda}'^* = \bar{\kappa}^* = 0 \,,  \quad \bar{v}^* = \frac{4-d}{2K_d} \,,  \label{eq:GaussFP}
\end{equation}
where the critical exponents assume their mean-field values \cite{Lawrie_1979}. The eigenvalues of the stability matrix of the Gaussian fixed point \eqref{eq:GaussFP} yield $1/\nu_t=2$ for largest eigenvalue, and 
$\phi_t = (4-d)/2$ for the ratio of the two relevant eigenvalues. 

If the specific-heat exponent $\alpha$ is positive, the susceptibility $\chi_{\varepsilon,k}$ of the energy-like density is expected to behave as $\chi_{\varepsilon,k} \sim k^{-\alpha/\nu}$ at the FRG fixed point \cite{Mesterhazy:2013naa}. As such, $\alpha/\nu$ can be obtained from the logarithmic $k$-derivative of $\chi_{\varepsilon,k}$ given by \eqref{eq:chiFlow},
\begin{equation}
    \frac{\alpha}{\nu} = -\frac{k\partial_k \chi_{\varepsilon,k}}{\chi_{\varepsilon,k}} = \frac{2K_d \bar{v}^*}{(1+\bar{m}^{2*})^3} \,.
\end{equation}
In the second equality, we have inserted the dimensionless couplings. At the Gaussian fixed point \eqref{eq:GaussFP}, 
this yields $\alpha_t/\nu_t = 4-d$.

Note that for $d>3$ the results for $\alpha_t/\nu_t= 4-d$ and $\phi_t= (4-d)/2$ appear to be inconsistent with the expected failure of hyperscaling relations above the upper critical dimension. This is a known subtlety of the RG \cite{Lawrie_1979}. The reason is that for $d>3$ the $(\phi^2)^3$ coupling $\kappa$ becomes dangerously irrelevant, which means that even though it is irrelevant by power counting, it is needed for stability of the effective potential, and can thus not be set  to zero directly \cite{10.1007/3-540-12675-9_11,Berche:2022wqt}. 
Although, strictly speaking, these results are therefore not valid for $d>3$ the advantage of keeping the number of dimensions $d$ in these hyperscaling relations for $d>3$, beyond their applicability, is that it allows for a direct comparison with results from the perturbative expansion in $\epsilon=4-d$ around four dimensions, with $\epsilon=0^+$, which has frequently been used in the past to study the tricritical dynamics of $\mathrm{^3He}$–$\mathrm{^4He}$ mixtures in the $N=2$ case as well \cite{PhysRevB.15.1427,Onuki1983,FolkMoser2007}.

\subsection{Dynamics}
\label{sec:resultsDynamics}

During the FRG flow, the finite FRG scale $k$ introduces an infrared cutoff on the correlation length, at $\xi \sim 1/k$, which is removed only in the $k\to 0$ limit. Hence, at a fixed point, the critical exponents of the kinetic coefficients can be read off from their power-law behavior with the FRG scale, i.e., $\Gamma_k^{\phi} \sim k^{-x_{\Gamma^{\phi}}}$ and $\gamma_k \sim k^{-x_{\gamma}}$. Since $k\,\partial_k \mu_k = 0$, the corresponding exponent is $x_{\mu} = 0$. 

Similarly, the dynamic critical exponent $z$ of a given mode can be deduced from the $k$-dependence of the lowest lying pole $\omega_k(p)$ in the dynamic response functions (with $p\equiv |\vec{p}|$ in isotropic systems). Modes with $p\ll k$ are non-critical, and thus the $p$-dependence of $\omega_k(p)$ is regular. For $k \ll p \ll \Lambda$ with $\Lambda \sim \pi T$ in hot QCD matter, the critical power law is manifest, $\omega_k(p) \sim p^z$. Requiring these two regimes to match smoothly at $p \sim k$, we find $\omega_k(k) \sim k^z$.

In our truncation, an explicit evaluation of the poles of the retarded propagator yields $\omega_{\phi,k}(p) = -i\Gamma_k^{\phi} (m_k^2+p^2)$, $\omega_{n,k}(p) = -i\gamma_k p^2/\chi_n$, and $\omega_{\varepsilon,k}(p) = -i\mu p^2/\chi_{\varepsilon,k}$. Setting $p \sim k$ and $m_k^2 = 0$ at the Gaussian fixed point, we find the dynamic critical exponents
\begin{equation}
    z_{\phi} = 2-x_{\Gamma^{\phi}}\,, \quad z_n=2-x_{\gamma}\,, \quad z_{\varepsilon}=2+\tilde{\alpha}/\nu \,, \label{eq:defsdynz} 
\end{equation}
where we have used $\chi_{\varepsilon,k} \sim k^{-\tilde{\alpha}/\nu}$ and the notation $\tilde{\alpha} \equiv \max\{\alpha,0\}$. The dynamic critical exponent $z_{\varepsilon}$ of the energy-like density is thus fully determined by static critical exponents (since $x_{\mu} = 0$). 

The fixed points of the FRG flow of the kinetic coefficients  are most conveniently analyzed using the following dimensionless ratios of relaxation times near criticality,
\begin{align}
    w_{n} \equiv \chi_n\frac{\Gamma_k^{\phi}}{\gamma_k} \,, \quad w_{\varepsilon} \equiv \chi_{\varepsilon,k} \frac{\Gamma_k^{\phi}}{\mu} \,, \label{eq:defws}
\end{align}
and the dimensionless effective coupling between the order parameter and the $O(N)$ charge densities,
\begin{align}
    f \equiv \frac{dK_d g^2 T k^{d-4}}{\Gamma_k^{\phi}\gamma_k} \,. \label{eq:fDef}
\end{align}
Using Eqs.~\eqref{eq:dGammaPhidk}, \eqref{Eq:dgammadk} and \eqref{Eq:sigma_kFlow}, the FRG flow equations of these ratios can be expressed as 
\begin{align} 
    k\partial_k w_{n} &= -w_{n}( x_{\Gamma^{\phi}} - x_{\gamma} ) \,,  \label{eq:w1Flow} \\[4pt] 
    k\partial_k w_{\varepsilon} &= -w_{\varepsilon}(x_{\Gamma^{\phi}}+\tilde{\alpha}/\nu) \,, \label{eq:w2Flow} \\[4pt] 
    k\partial_k f &= -f(4-d-x_{\Gamma^{\phi}}-x_{\gamma}) \, ,\label{eq:f1Flow}
\end{align}
with
\begin{align}
    x_{\Gamma^{\phi}} \equiv -\frac{k\partial_k \Gamma^{\phi}_k}{\Gamma^{\phi}_k} \,, \quad x_{\gamma} \equiv -\frac{k\partial_k \gamma_k}{\gamma_k} \,.
\end{align}
At the fixed points of the FRG flow these become scale $k$ independent, implying the power laws $\Gamma_k^{\phi} \sim k^{-x_{\Gamma^{\phi}}}$ and $\gamma_k \sim k^{-x_{\gamma}}$ in the dynamic scaling regime.

Moreover, when the scale derivatives on the left sides in Eqs.~\eqref{eq:w1Flow}, \eqref{eq:w2Flow} and \eqref{eq:f1Flow}
vanish, either the corresponding finite fixed-point values of $w_n$, $w_\varepsilon $ and $f$ themselves vanish, or else we obtain a dynamic scaling relation for the corresponding scaling exponents from the terms in brackets for each non-vanishing one, as discussed below.

The decomposition \eqref{eq:dGammaPhidk} entails that $x_{\Gamma^{\phi}}$ can be conveniently separated into contributions from the reversible coupling to the $O(N)$ charge densities, and from the dissipative coupling to the energy-like density,\footnote{More specifically, $x_{\Gamma^{\phi}}^{(n)}$ corresponds to the contributions from the first two diagrams and $x_{\Gamma^{\phi}}^{(\varepsilon)}$ to that from the third diagram in the first line of Fig.~\ref{fig:flowOf2PtFnc}, as explained in the figure caption.}
\begin{align}
    x_{\Gamma^{\phi}} = x_{\Gamma^{\phi}}^{(n)} + x_{\Gamma^{\phi}}^{(\varepsilon)} \, , \label{eq:xGammaPhi}
\end{align}
in which the individual contributions from $n$ and $\varepsilon$ are given explicitly by
\begin{widetext}
\begin{align}
    x_{\Gamma^{\phi}}^{(n)} &= \frac{(N-1)f}{(d-2) (1+\bar{m}^2)^2} \left((d-4) \, _2F_1\left(1,\frac{d}{2}-1;\frac{d}{2};-\frac{1}{(1+\bar{m}^2)w_n}\right)+\frac{2+(4-d)
   (1+\bar{m}^2) w_n}{(1+\bar{m}^2) w_n+1}\right)\, , \label{eq:xGammaPhiN} \\
    x_{\Gamma^{\phi}}^{(\varepsilon)} &= -\frac{dK_d \bar{v}}{N(1+\bar{m}^2)^4} \bigg(-\frac{d}{(d+2) w_{\varepsilon }} \, _2F_1\left(1,\frac{d}{2}+1;\frac{d}{2}+2;-\frac{1}{(1+\bar{m}^2)w_{\varepsilon }}\right)\; +  \label{eq:xGammaPhiE} \\ \nonumber
    &\hspace{2cm} 2
   (1+\bar{m}^2) \left(\left(\frac{2}{d}-1\right) \, _2F_1\left(1,\frac{d}{2};\frac{d}{2}+1;-\frac{1}{(1+\bar{m}^2)w_{\varepsilon }}\right)+\frac{(1+\bar{m}^2)
   w_{\varepsilon }}{(1+\bar{m}^2)w_{\varepsilon }+1}\right)+\frac{1+\bar{m}^2}{(1+\bar{m}^2) w_{\varepsilon }+1}\bigg) \, .
\end{align}
\end{widetext}

On the other hand, the FRG flow of $\gamma_k$, which determines that of the charge diffusion and is represented by the logarithmic $k$-derivative $x_{\gamma}$, receives contributions only from the reversible coupling to the order parameter, and is simply given by 
\begin{align}
    x_{\gamma} = \frac{2f}{d(1+\bar{m}^{2})^3} \label{eq:xgamma}
\end{align}

We now restrict our discussion to the Gaussian fixed point \eqref{eq:GaussFP}. 
In general, the positive specific-heat exponent $\alpha_t = (4-d)/2$ and the non-renormalization of $\mu$ entail that the energy-like density has a dynamic critical exponent of 
\begin{equation}
    z_{\varepsilon} = 2+\frac{\alpha_t}{\nu_t} 
\end{equation}
which amounts to $z_{\varepsilon} = 3$ with $\alpha_t/\nu_t =1 $ in the relevant case of $d=3$ spatial dimensions. More specifically, its diffusion coefficient $D_{\varepsilon} = \mu/\chi_{\varepsilon}$ vanishes as $D_{\varepsilon} \sim k^{\alpha_t/\nu_t}$.
\begin{figure}[t]
    \centering
    \textsf{\large \hspace{1.0cm} $N=4$ (QCD)}\\
    \includegraphics[width=0.8\linewidth]{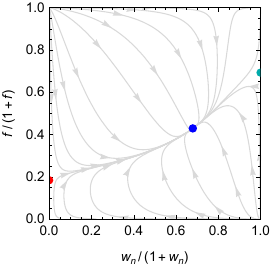}
    \caption{FRG flow in the compactified $(w_n,f)$ plane for $N=4$, $d=3$, at the Gaussian fixed point \eqref{eq:GaussFP} and with $w_{\varepsilon} \to \infty$.}
    \label{fig:flowDynamicN4}
\end{figure}
To analyze how the energy-like density affects the critical dynamics of the order parameter, we first consider the flow equation \eqref{eq:w2Flow} of $w_{\varepsilon}$. One can show that $x_{\Gamma^{\phi}} > -2(4-d)/N$ (see Appendix~\ref{app:ineq}), and therefore $k\partial_k w_{\varepsilon}$ satisfies the inequality
\begin{align}
    k\partial_k w_{\varepsilon} 
    < -w_{\varepsilon}(4-d)\left(1-\frac{2}{N}\right) \,.
\end{align}
Hence, $k\partial_k w_{\varepsilon}$ is strictly negative for $N \geq 2$, which implies that $w_{\varepsilon}$ strictly increases for $k\to 0$ and therefore the fixed point is at $w_{\varepsilon}^* = \infty$
(in fact, \eqref{eq:w2Flow} then simply entails that $w_\varepsilon \sim k^{-(x_{\Gamma_\phi}+\tilde\alpha/\nu)}$ which immediately follows from its definition in \eqref{eq:defws}). 

\begin{figure}[!t]
    \centering
    \textsf{\large \hspace{1.0cm} $N=2$ ($\mathsf{^3He}$–$\mathsf{^4He}$ mixtures)}\\
    \includegraphics[width=0.8\linewidth]{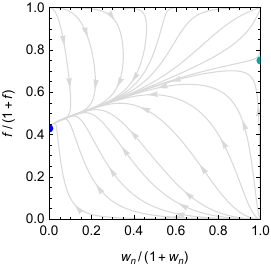}
    \caption{FRG flow in the compactified $(w_n,f)$ plane for $N=2$, $d=3$, at the Gaussian fixed point \eqref{eq:GaussFP} and with $w_{\varepsilon} \to \infty$.}
    \label{fig:flowDynamicN2}
\end{figure}

In the remaining $(w_n,f)$-plane, the FRG flow is shown (with $d=3$) in Figure \ref{fig:flowDynamicN4} for $N=4$ and in Figure \ref{fig:flowDynamicN2} for $N=2$.
As one can see, there are two possible scenarios, depending on the values of $N$ (and $d$). For $N$ sufficiently large, cf.~Fig.~\ref{fig:flowDynamicN4}, the stable fixed point is located at $f^*=(4-d)d/4$ and a value of $w_n$ that is indirectly determined by
\begin{align}
    _2F_1\left(1,\frac{d}{2}-1;\frac{d}{2};-\frac{1}{w_n^*}\right) &= \label{eq:wnFPEq} \\ \nonumber
    &\hspace{-2.0cm} \frac{1}{4-d}\bigg[ \frac{2+(4-d)w_n^*}{1+w_n^*} - \frac{2(d-2)(N+4)}{d\,N(N-1)} \bigg] \,. 
\end{align}
In the relevant case of $d=3$ spatial dimensions, this equation has a non-vanishing and finite ($0<w_n^*<\infty$) solution only if $N>2$. This includes, in particular, the $N=4$ case relevant for QCD, where we numerically find $w_n^* \approx 2.119$.  The finiteness of $w_n^*$ and $f^*$ in \eqref{eq:w1Flow} and \eqref{eq:f1Flow} then imply the \emph{strong dynamic scaling} relation 
\[ x_{\Gamma^{\phi}}=x_{\gamma}=2-d/2 \, , \]
at the Gaussian fixed point \eqref{eq:GaussFP}, i.e., the tricritical point in $d=3$. It implies that the fluctuations of the order parameter and the fluctuations of the $O(N)$ charge densities both relax on the same characteristic timescale. From \eqref{eq:defsdynz} the dynamic critical exponents are
\begin{align}
    z_{\phi} = z_{n} = \frac{d}{2} \, , \label{eq:zStrongScaling}
\end{align}
as in $O(N)$ Model G \cite{Rajagopal:1992qz} a.k.a.~the SSS model \cite{SASVARI1975108,taeuber_2014}. Here, however, we
have the peculiar situation that $z_\varepsilon > z_\phi=z_n$, i.e., the fluctuations of the order parameter and the $O(N)$ charge densities both relax on a timescale of order $k^{-d/2}$, which for $k\to 0$ becomes much shorter, however, than the timescale of order $k^{-(2+\alpha_t/\nu_t)}$ over which the fluctuations of the energy-like density diffuse. Physically, this akin to the region of `anomalous diffusion' in the phase diagram of Model~C \cite{Mesterhazy:2013naa}. 

\begin{figure}
    \centering
    \includegraphics[width=0.8\linewidth]{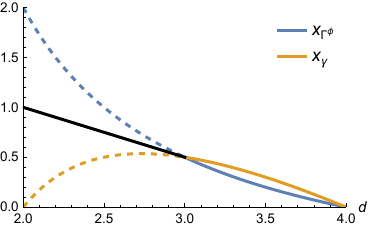}
    \caption{Critical exponents \eqref{eq:wfFP1CritExp} for $N=2$ as a function of $d$ in the range $2<d<4$. For $d<3$ the  fixed-point structure is qualitatively the same as in Fig.~\ref{fig:flowDynamicN4}. The solid black line shows the strong-scaling values $x_{\Gamma^{\phi}}=x_{\gamma}=2-d/2$, and the dashed lines show the values of the critical exponents at the (unstable) weak-scaling fixed point with $w_n=0$. In the limit $d\to 3^-$ the strong-scaling fixed point merges with the weak-scaling fixed point, leading to the (stable) weak-scaling behavior for $d>3$. For an extension of this plot to $N\neq 2$, we refer to Appendix~\ref{sec:pd}.} 
    \label{fig:FPMergerN2}
\end{figure}

If the values of $N$ and $d$ are such that Eq.~\eqref{eq:wnFPEq} has no non-trivial solution for $w_n^*$, the stable fixed point is the weak-scaling fixed point at
\begin{align}
    w_n^* = 0 \,, \quad f^* = \frac{(4-d)(d-2)d(N+2)}{2N(dN-2)} \label{eq:wfFP1}
\end{align}
as illustrated in Fig.~\ref{fig:flowDynamicN2}. In $d=3$ spatial dimensions, this happens for $N \leq 2$. In particular, the limiting value $N=2$ is the relevant case for the tricritical point in $\mathrm{^3He}$--$\mathrm{^4He}$ mixtures. Since $w_n^* = 0$, the strong-scaling relation \eqref{eq:zStrongScaling} is no longer guaranteed, but one is left with the weaker dynamic scaling relation
\begin{equation}
    x_{\Gamma^{\phi}}+x_{\gamma}=4-d \,, \label{eq:scalRel}
\end{equation}
which is implied by  \eqref{eq:f1Flow} alone, using the finiteness of $f^*$. Hence, $x_{\Gamma^{\phi}}$ and $x_{\gamma}$ can assume different values. We now restrict our discussion to the $N=2$ case. At the weak-scaling fixed point \eqref{eq:wfFP1}, the critical exponents \eqref{eq:xGammaPhi} and \eqref{eq:xgamma} are explicitly given by
\begin{equation}
    x_{\Gamma^{\phi}} = \frac{3}{d-1}-1\,, \quad x_{\gamma}=5-d-\frac{3}{d-1} \,. \label{eq:wfFP1CritExp}
\end{equation} 
Expanding to first order $\epsilon=4-d$, we find
\begin{align}
    x_{\Gamma^{\phi}} &= 
    \frac{\epsilon}{3}+\mathcal{O}(\epsilon^2) \,, \quad   
    x_{\gamma} = \frac{2\epsilon}{3}+\mathcal{O}(\epsilon^2) \,, \label{eq:tricritExpSN}
\end{align}
in agreement with Model~F' \cite{PhysRevB.15.1427,Onuki1983}. Remarkably, however, directly setting $d=3$ in \eqref{eq:wfFP1CritExp} yields $x_{\Gamma^{\phi}} = 2-d/2=x_{\gamma}$, resembling \emph{strong} scaling even though the latter is not guaranteed here. 
The reason is that, for $N=2$ fixed, a strong-scaling fixed point  appears just below $d=3$. 
In the limit $d\to 3^-$, this strong-scaling fixed point merges with the weak-scaling fixed point at $w_{n} = 0$. 
This behavior is illustrated in Fig.~\ref{fig:FPMergerN2}, where the critical exponents at the stable fixed point are shown as a function of $d$ between $2<d<4$. One can see that the differing values of $x_{\Gamma^{\phi}} \neq x_{\gamma}$ in the weak-scaling regime $d>3$ merge at $d=3$. For $d<3$ the stable strong-scaling fixed point appears, and the critical exponents are given by \eqref{eq:zStrongScaling}, shown as the solid black line in the figure. As such, it is not surprising that the strong-scaling behavior persists in exactly $d=3$ spatial dimensions, when regarded as the limit $d \to 3^-$.

We emphasize, however, that while Fig.~\ref{fig:FPMergerN2} is certainly useful for comparisons with the $\epsilon$-expansion and to illustrate the transition from strong-scaling to weak-scaling behavior, respectively, the 
values of the critical exponents are unconditionally valid only at the point $d=3$. This is because, for $d>3$, the expected failure of hyperscaling relations requires additional care in  the interpretation of critical exponents, as briefly discussed in the context of dangerously irrelevant couplings at the end of Sec.~\ref{sec:resultsStatics}. On the other hand, for $d<3$, i.e.~below the upper critical dimension of three, the Gaussian fixed point no-longer describes the tricritical point. (This is analogous to ordinary critical points, where the Wilson-Fisher fixed point takes over the role of the Gaussian fixed point below the upper critical dimension of four.) Addressing these issues by the corresponding proper extensions to $d\neq 3$ 
is beyond the scope of our present work and will not affect the physically most relevant case of $d=3$ spatial dimensions. 

In summary, we find that strong scaling $z_{\phi}=z_{n}=d/2$ is preserved in the presence of the energy-like density, the latter becoming \emph{subdiffusive} with $z_{\varepsilon}=2+\alpha_t/\nu_t$. With due appreciation of possible systematic errors induced by our truncation, we find strong scaling for all $N \geq 2$, covering \emph{both} the tricritical point of QCD and the tricritical point in $\mathrm{^3He}$--$\mathrm{^4He}$ mixtures where the latter is special as an interesting limiting case of strong scaling for $d\to 3^-$ and $N\to 2^+$ (see Appendix~\ref{sec:pd}).

\subsection{Influence of shear modes}
\label{eq:influenceOfShearModes}

In this section, we comment on the influence of the diffusive shear modes, 
which we have neglected. First of all, the advection of a non-conserved order parameter by shear modes can be argued to be RG irrelevant \cite{Hohenberg:1977ym}. As such, the shear modes should not affect the tricritical dynamics of the order parameter. However, they might affect the (heat) conductivity $\mu$ of the energy-like density, and its associated critical exponent $x_{\mu}$ which vanishes in their absence, cf.~Eq.~\eqref{Eq:sigma_kFlow}. 
In the following, we argue that the influence of shear modes can be qualitatively understood by adapting the mode-coupling argument by Halperin and Hohenberg \cite{Hohenberg:1977ym} (who attribute it to Arcovito et al.~\cite{PhysRevLett.22.1040}) to the tricritical point.

The conductivity $\mu$ can generally be measured by applying an external `electric' field $\vec{E}$ which couples to the energy-like density. Without advection by the fluctuating momentum density, this field would induce a finite current $\vec{j}_{\varepsilon} = \mu \vec{E}$, and thus $x_{\mu} = 0$. 
With advection, on the other hand, a chunk of fluid of linear extent $\xi$ will experience an applied net force of the order $\vec{f}_{\text{app}} \sim \varepsilon\xi^d \vec{E}$, and will accelerate until the force due to viscous drag $\vec{f}_{\text{visc}} \sim -\bar{\eta} \vec{v} \xi^{d-2}$ balances the applied force, where $\bar{\eta}$ denotes the shear viscosity. The resulting current is
\begin{equation}
    \vec{j}_{\varepsilon} = \varepsilon\vec{v} \sim \frac{\varepsilon^2 \xi^2}{\bar{\eta}} \vec{E} \,.
\end{equation}
Its expectation value is determined by the equipartition theorem $\langle(\delta \varepsilon)^2 \rangle = T\chi_{\varepsilon} /\xi^d$.  As discussed in Sec.~\ref{sec:resultsStatics} above, the susceptibility $\chi_{\varepsilon}$ diverges as $\chi_{\varepsilon} \sim \xi^{\alpha_t/\nu_t}$.
 We thus obtain the following contribution to $\mu$ from fluctuations on the scale of $\xi$,
\begin{equation}
    \mu \sim (1/\bar{\eta})\, \xi^{2-d+\alpha_t/\nu_t} \,.  \label{eq:critContribToLambdan}
\end{equation}
With $ \alpha_t/\nu_t = 1$ in the relevant case of $d=3$ dimensions the exponent on the right-hand side of \eqref{eq:critContribToLambdan} vanishes, suggesting that there is no power-law contribution from tricritical fluctuations. 
By this argument, one would expect that $x_{\mu} = 0$ at the tricritical point, even with the shear modes included which should thus not affect our analysis of tricritical dynamic scaling presented above. It would be insightful to compare this result to the tricritical point in ${}^3\mathrm{He}$--${}^4\mathrm{He}$ mixtures. We are not aware, however, of any work that studies the influence of shear modes on the tricritical point in Model~F'.

\section{Explicit symmetry breaking}
\label{sec:explSymmBreak}

Non-vanishing current quark masses in QCD explicitly break the $O(4)$ chiral symmetry, turning the second-order chiral phase transition into an analytic crossover and the tricritical point into an ordinary $Z_2$ critical point, see Fig.~\ref{fig:pdiag5}. The dynamic universality class of the $Z_2$ critical point is plausibly that of Model~H, i.e.,~that of the liquid-gas critical point \cite{Son:2004iv}. It is interesting to see how the tricritical dynamics from the previous section re-organizes when the $O(N)$ symmetry is explicitly broken. In the following, we identify the corresponding dynamic universality class by studying the eigenmodes of the linearized equations of motion around the mean-field solution. In particular, at the upper critical dimension ($d=3$) the mean-field approximation is sufficient to reproduce tricritical scaling up to possible logarithmic corrections.

\subsection{\boldmath Static $Z_2$ criticality  near the tricritical point}

We introduce an external symmetry-breaking field $H$ in the LGW functional \eqref{eq:freeEnergy}, which (with $\rho \equiv \phi^2$) thus then becomes
\begin{align}
    F &= \int d^dx \, \bigg\{ \frac{1}{2}(\vec{\nabla}\phi)^2 + U(\rho,\varepsilon) + \frac{n_{ab} n_{ab}}{4\chi_n} - H\sigma \bigg\} \, , \label{eq:freeEnergyWithH}
\end{align}
with the potential
\begin{align}
    U(\rho,\varepsilon) = \frac{m^2}{2}\rho + \frac{\lambda}{4!N} \rho^2 &+ \frac{\kappa}{6!N^2} \rho^3 \label{eq:effPotMF}\\
    &+ \frac{\iota}{2\sqrt{N}} \rho \varepsilon + \frac{\varepsilon^2}{2\chi_{\varepsilon}} \, , \nonumber
\end{align}
where we have chosen coordinates such that the condensate is oriented along the $\sigma \equiv \phi_{N}$ direction in the $N$-dimensional field space. In mean-field approximation, the equilibrium state is given by the global minimum of the potential \eqref{eq:effPotMF}.
First, we minimize the potential $U(\rho,\varepsilon)$ with respect to $\varepsilon$ while keeping $\rho$ fixed, which yields the $\rho$-dependent minimum value of $\varepsilon$,
\begin{align}
    \varepsilon_0(\rho) = -\frac{\iota \chi_{\varepsilon}}{2\sqrt{N}} \rho \, .
\end{align}
Inserting this value back into the potential \eqref{eq:effPotMF} yields
\begin{align}
    V(\rho) &= \min_{\varepsilon} U(\rho,\varepsilon) \\
    &=\frac{m^2}{2}\rho + \frac{\lambda'}{4!N} \rho^2 + \frac{\kappa}{6!N^2} \rho^3 \, , \label{eq:effPotMFRho}
\end{align}
with the shifted quartic coupling $\lambda' = \lambda-3\iota^2\chi_{\varepsilon}$ introduced as in Sec.~\ref{sec:rFRGA} above. Secondly, in presence of the external field $H$, the global minimum of the tilted potential is then determined by the condition
\begin{align}
    2\sigma_0 V'(\rho_0) = H \,. \label{eq:groundStateCond}
\end{align}
The $Z_2$ critical point occurs if both the second and third derivatives of $V(\rho)$ with respect to $\sigma$ also vanish,
\begin{align}
    0&= \frac{\partial^2 V}{\partial \sigma^2}\bigg\vert_{\sigma_c}  = 2V'(\sigma_c^2) + 4\sigma_c^2 V''(\sigma_c^2) \, , \\
    0&= \frac{\partial^3 V}{\partial \sigma^3}\bigg\vert_{\sigma_c} = 4\sigma_c\left(3V''(\sigma_c^2) + 2\sigma_c^2 V''(\sigma_c^2) \right)\, .
\end{align}
For the potential \eqref{eq:effPotMFRho} expanded up to third order in $\rho$, this yields the conditions
\begin{align}
    H &= m^2 \sigma_c + \frac{\lambda'}{6N}\sigma_c^3 + \frac{\kappa}{5!N^2} \sigma_c^5 \, ,\\
    0 &= m^2 + \frac{\lambda'}{2N}\sigma_c^2 + \frac{\kappa}{4!N^2} \sigma_c^4 \, , \\
    0 &= \frac{\lambda'}{N}\sigma_c + \frac{\kappa}{6N^2} \sigma_c^3\, ,
\end{align}
which are solved by \cite{Pradeep:2019ccv}
\begin{align}
    \sigma_c = \left(\frac{45 N^2 H}{\kappa}\right)^{1/5} \!\!, \quad
    m^2_c = \frac{\kappa \sigma_c^4}{24 N^2} \,, \quad \lambda'_c = -\frac{\kappa \sigma_c^2}{6N} \, .
\end{align}
In particular, this implies 
$\sigma_c\sim H^{1/5}$, $m_c^2 \sim H^{4/5}$ and $\lambda'_c \sim H^{2/5}$.

From the distribution $e^{-F/T}$ one can derive the two-point correlation functions of $\sigma$, $\vec{\pi}$ and $\varepsilon$ in the mean-field approximation (where $V$ now denotes the volume of the system),
\begin{align}
    V\langle (\delta\sigma)^2 \rangle_{\vec{p}} &= \frac{T}{m_{\sigma}^2+\vec{p}^2} \,, \\
    V\langle (\delta\pi_i)^2 \rangle_{\vec{p}} &= \frac{T}{m_{\pi}^2+\vec{p}^2} \,, \\
    V\langle (\delta \varepsilon)^2\rangle_{\vec{p}} &=
    T\chi_{\varepsilon} + \frac{T \sigma_0^2 \iota^2 \chi_{\varepsilon}^2}{N m_{\sigma}^2} \, .\label{eq:eVar}
\end{align}
Hence, the correlation length in the order parameter ($\sigma$) channel is given by $\xi=1/m_{\sigma}$. The screening mass $m_{\pi}$ of pions scales as \cite{Stephanov:1999zu}
\begin{align}
    m_{\pi}^2 = \frac{H}{\sigma_c} \sim H^{4/5} .\label{eq:mPi2}
\end{align}
The static fluctuations of pions become negligible for $\xi \gg m_{\pi}^{-1}$. In other words, in order for pure $Z_2$ scaling to remain, $\xi$ must be at least of order $H^{-2/5}$. As argued in Ref.~\cite{Son:2004iv}, the estimate for the typical correlation length $\xi\sim 2-3$~fm reached in a heavy-ion collision is not much larger than the Compton wavelength of the pion $1/m_{\pi} \approx 1.4$~fm.

\subsection{Dynamics near the tricritical point}
\label{sec:VB}

At long wavelengths $p\xi \ll 1$ the equations of motion \eqref{eq:eoms} may be linearized around the mean field $\sigma=\sigma_0+\delta\sigma$, $\vec{\pi}=\delta\vec{\pi}$, $\varepsilon=\varepsilon_0+\delta\varepsilon$, $n_{ab}=\delta n_{ab}$.
First, the equations of motion for $\delta\sigma$ and $\delta\varepsilon$ are
\begin{align}
    \frac{\partial \delta\sigma}{\partial t} &= -\Gamma^{\phi}( (-\vec{\nabla}^2 + m_{\sigma}^2 + \frac{\iota^2 \chi_{\varepsilon} \sigma_0^2}{N} 
    ) \delta\sigma + \frac{\iota \sigma_0}{\sqrt{N}} \delta\varepsilon ) + \theta_0 \, ,\nonumber \\
    \frac{\partial \delta\varepsilon}{\partial t} &= \mu\vec{\nabla}^2 ( \frac{\iota \sigma_0}{\sqrt{N}} \delta\sigma + \frac{\delta \varepsilon}{\chi_{\varepsilon}}  ) + \Xi \, .\label{eq:linEoMSigmaAndEps}
\end{align}
For $H>0$ finite, we have $\sigma_0 > 0$, and therefore $\delta\sigma$ and $\delta\varepsilon$ mix. This leads to a different qualitative behavior of the eigenmodes, as first noted in Ref.~\cite{Son:2004iv}.
In matrix notation, the equations \eqref{eq:linEoMSigmaAndEps} can be written as
\begin{align}
    \frac{\partial}{\partial t}
    \begin{pmatrix}
        \delta\sigma\\ \delta\varepsilon
    \end{pmatrix} &= \\ \nonumber
    &\begin{pmatrix}
        -\Gamma^{\phi}( -\vec{\nabla}^2 + a) & -\Gamma^{\phi}\frac{\iota \sigma_0}{\sqrt{N}} \\
        \mu\vec{\nabla}^2 \frac{\iota \sigma_0}{\sqrt{N}} & \mu\vec{\nabla}^2/\chi_{\varepsilon}
    \end{pmatrix}
    \begin{pmatrix}
        \delta\sigma\\ \delta\varepsilon
    \end{pmatrix}+
    \begin{pmatrix}
        \theta_0 \\
        \Xi
    \end{pmatrix}
\end{align}
The eigenfrequencies are 
\begin{align}
    \omega_1 &= -i \Gamma^{\phi} \left(m_{\sigma}^2+\frac{\iota^2 \chi_{\varepsilon} \sigma_0^2}{N}\right) +\, O(\vec{p}^2) \,, \\
    \omega_2 &= -i D \vec{p}^2 \,+\, O(\vec{p}^4)\,, \label{eq:eigModes}
\end{align}
with diffusion coefficient
\begin{equation}
    D = \frac{\mu}{\chi_{\varepsilon} + \frac{\sigma_0^2 \iota^2 \chi_{\varepsilon}^2}{N m_{\sigma}^2}} \, .
\end{equation}
For $m_{\sigma}^2 \ll \frac{\iota^2 \chi_{\varepsilon} \sigma_0^2}{N}$ the frequencies become
\begin{align}
    \omega_1 &= -i \Gamma^{\phi} \frac{\iota^2 \chi_{\varepsilon} \sigma_0^2}{N} +O(\vec{p}^2) \,, \\
    \omega_2 &= -i \frac{\mu N m_{\sigma}^2}{\sigma_0^2 \iota^2 \chi_{\varepsilon}^2}  \vec{p}^2 +O(\vec{p}^4)\,, 
\end{align}
Hence, for $H>0$ fixed, the relaxation rate of the first mode stays finite as the $Z_2$ line is approached $m_{\sigma}^2\to 0$, while the diffusion coefficient of the second mode vanishes as $D \sim m_{\sigma}^2 \sim \xi^{-2} \to 0$.

Next, we consider the equations of motion of $\delta\pi_i$ and the iso-axial-vector charge densities $\delta n_{0i}$,
\begin{align}
    \frac{\partial \delta\pi_i}{\partial t} &= -\Gamma^{\phi} (-\vec{\nabla}^2 + m_{\pi}^2 
    ) \delta\pi_i + \frac{g\sigma_0}{\chi_n} \delta n_{0i} + \theta_i \, , \label{eq:linEoMPi} \\
    \frac{\partial \delta n_{0i}}{\partial t} &= \frac{\gamma}{\chi_n} \vec{\nabla}^2 \delta n_{0i} + g \sigma_0 (\vec{\nabla}^2-m_{\pi}^2) \delta \pi_i + \zeta_{0i} \, , \label{eq:linEoPN}
\end{align}
which are also coupled for $\sigma_0 \neq 0$. The equations of motion of the iso-vector charge densities remain diffusive,
\begin{align}
    \frac{\partial \delta n_{ij}}{\partial t} &= \frac{\gamma}{\chi_n} \vec{\nabla}^2 \delta n_{ij}  + \zeta_{ij} \, . 
\end{align}
In matrix notation, \eqref{eq:linEoMPi} and \eqref{eq:linEoPN} can be written as
\begin{align}
    \frac{\partial}{\partial t}
    \begin{pmatrix}
        \delta\pi_i\\ \delta n_{0i}
    \end{pmatrix} &= \\ \nonumber
    &\hspace{-0.5cm} \begin{pmatrix}
         -\Gamma^{\phi} (-\vec{\nabla}^2+m_{\pi}^2) &  g\sigma_0/\chi_n \\
        g \sigma_0 (\vec{\nabla}^2-m_{\pi}^2) & \gamma \vec{\nabla}^2/\chi_n 
    \end{pmatrix}
    \begin{pmatrix}
        \delta\pi_i\\ \delta n_{0i}
    \end{pmatrix} +
    \begin{pmatrix}
        \theta_i \\
        \zeta_{0i}
    \end{pmatrix} \, .
\end{align}

Provided that the damping is sufficiently small, the two eigenmodes are both propagating, with frequencies \cite{Son:2002ci,Florio:2021jlx}
\begin{align}\label{Eq:omega+-}
    \omega_{\pm}(\vec{p}) = \pm \nu(\vec{p})-\frac{i}{2}\Gamma_{\pi}(\vec{p})  \,,
\end{align}
and
\begin{align}
    \nu^2(\vec{p}) &\approx v^2(m^2+\vec{p}^2)  \,, \label{eq:nuWeakDamping} \\
    \Gamma_{\pi}(\vec{p}) &\approx \Gamma^{\phi}m_{\pi}^2 + (\Gamma^{\phi}+\frac{\gamma}{\chi_n}) \vec{p}^2 \,,
\end{align}
where $v = g\sigma_0/\sqrt{\chi_n}$ denotes the pion velocity. In this case, relaxation to equilibrium is accompanied by oscillatory behavior with oscillation frequency given by $\nu$.

The pion pole mass is given by $m_{\pi}^p = \nu(0) = vm_{\pi}$. To see how $m_{\pi}^p$ behaves along the $Z_2$ line, recall that $\sigma_c^2 \sim H^{2/5}$, $m_{\pi}^2 \sim H^{4/5}$, and that $g$ and $\chi_n$ do not depend on $H$, which entails
\begin{equation}
    m_{\pi}^p \sim m_{\pi}\sigma_0 \sim H^{3/5} \,. \label{eq:pionPoleMassScaling}
\end{equation}
This can be compared to the pions in vacuum, where the scaling of the pion pole mass is determined by the Gell-Mann--Oakes--Renner relation, $m_{\pi}^p \sim H^{1/2}$, which is parametrically larger than \eqref{eq:pionPoleMassScaling} for sufficiently small $H$. In other words, this argument would imply that the pion pole mass is reduced at the $Z_2$ critical point, if the latter is sufficiently close to the tricritical point.
The result \eqref{eq:pionPoleMassScaling} can also be compared with the $O(4)$ line, where one would expect the critical scaling $m_{\pi}^p \sim H^{(1+1/\delta)/2}$ \cite{Son:2001ff}. Using $\delta = 4.824(9)$ from the $O(4)$ universality class \cite{Engels:2003nq}, this would approximately amount to $m_{\pi}^p \sim H^{0.6036}$, which is very close to $m_{\pi}^p \sim H^{0.6}$ from \eqref{eq:pionPoleMassScaling}. 

In summary, after a time of the order $1/m_{\pi}^p \sim H^{-3/5}$ the only remaining mode \eqref{eq:eigModes} is diffusive,\footnote{Near the $Z_2$ line, the timescale of the critical mode is determined by $\xi_t \sim \xi^{z_{\phi}}$ with $z_\phi \approx 3$. Hence, $\xi$ must at least be of the order $H^{-1/5}$ for the pions to become irrelevant to the dynamics, as compared to the static constraint of  $\xi $ being  larger than  $H^{-2/5}$ from Eq.~\eqref{eq:mPi2}. } and the corresponding diffusion coefficient scales as $D \sim \xi^{-2}$ near the $Z_2$ line. This is in contrast to the tricritical scaling in the chiral limit, where we found $D_\varepsilon \sim \xi^{-\alpha_t/\nu_t} \sim \xi^{-1}$, as discussed in Sec.~\ref{sec:resultsDynamics}. As we have seen in Eqs.~\eqref{eq:linEoMSigmaAndEps} above, the reason for this difference is that the energy-like density $\varepsilon$ mixes linearly with the chiral order parameter $\sigma$ for $H \neq 0$ and hence $\sigma_0 \neq 0 $, when chiral symmetry is explicitly broken. This then leads to
the dynamic universality class of Model~B \cite{Hohenberg:1977ym}, here in our mean-field approximation. Beyond mean-field, the diffusion coefficient of the conserved charge in Model~B, whose role is here being played by the slow mode \eqref{eq:eigModes} in the mixing of $\delta \sigma $ and  the energy-like density $\delta\varepsilon$, vanishes as $D \sim \xi^{-2+\eta}$, where $\eta \approx 0.036$ denotes the small (static) anomalous dimension of the $Z_2$ order-parameter field.

In this situation, on the other hand, when the dynamics is no-longer dominated by the tricritical point, near the $Z_2$ critical point the fluctuations of the momentum density need  to be included as well. The dynamic universality class is then expected to turn into that of Model~H \cite{Son:2004iv}, where the diffusion coefficient scales as $D \sim \xi^{-2+\eta+x_{\mu}}$. Using the dynamic scaling relation $x_{\mu}+x_{\eta} = 4-d-\eta$ of Model~H, the exponent can be written as $D \sim \xi^{-(1+x_{\eta})}$ in $d=3$ spatial dimensions. Since $x_{\eta} \approx 0.05$ is well-known to be small \cite{10.1143/PTP.55.1384,Roth:2024hcu}, this is then quite close to the result $D_\varepsilon\sim \xi^{-1}$, i.e.~practically indistinguishable from tricritical scaling. In other words, the subdiffusive fluctuations of the energy-like density $\varepsilon$, with dynamic critical exponent $z_\varepsilon = 3$ near the tricritical point, turn into those of the conserved order parameter, i.e.~the specific entropy with $z = 3 + x_\eta$, as the mixing increases with increasing distance from the tricritical point close to the $Z_2$ line. Which one is realized in QCD really depends on the time scales and the proximity of the tricritical point.      

\section{Summary and Outlook}
\label{sec:summaryAndOutlook}

In the limit of two massless quark flavors, the chiral phase transition of QCD is expected to be second order and to fall into the $O(4)$ universality class \cite{Pisarski:1983ms}. 
The dynamic universality class is plausibly the one of Model~G from the Halperin-Hohenberg classification \cite{Hohenberg:1977ym} (i.e., the one of a Heisenberg antiferromagnet) generalized to a four-component order-parameter \cite{Rajagopal:1992qz}.

At non-zero baryon chemical potential,
it is conjectured that the $O(4)$ line ends in a tricritical point \cite{Halasz:1998qr} beyond which the transition becomes first order.  The main difference between tricritical scaling and that along the $O(4)$ line is that the triciritical specific-heat exponent $\alpha_t=1/2$ is positive while it is negative for all $O(N)$ with $N\ge 2$ (in $d=3$ dimensions). Therefore, the fluctuating \mbox{energy(-like)} density might affect the critical dynamics. 
To study the tricritical dynamics, we have supplemented the equations of motion of the S\'asvari-Schwabl-Sz\'epfalusy (SSS) model \cite{SASVARI1975108} (i.e., the generalization of Model~G to an $N$-component order parameter) with a conserved energy-like density. Following the naming convention of Ref.~\cite{Folk_2006} for Model F', this model might analogously be referred to as the \emph{SSS' model}, where the prime indicates the coupling to an additional conserved density.

We have studied the static critical behavior of this model with the functional renormalization group (FRG) \cite{Wetterich:1992yh}. In $d=3$ spatial dimensions, the tricritical point corresponds to a Gaussian fixed point of the FRG flow, for which we have reproduced the known mean-field exponents $\nu_t=1/2$ and $\alpha_t=1/2$ of the corresponding $\phi^6$-theory. 

Using the real-time formulation of the FRG developed recently \cite{Roth:2024rbi,Roth:2024hcu}, we have derived non-perturbative flow equations for the kinetic coefficients. We have shown that for $N \geq 2$ the order parameter $\phi_a$ and the $N(N-1)/2$ charge densities $n_{ab}$ satisfy the strong-scaling relation $z_{\phi}=z_{n}=3/2$ near the tricritical point, as they do in the original SSS model \cite{taeuber_2014}. In contrast, the energy-like density becomes \emph{subdiffusive} with the larger dynamic critical exponent $z_{\varepsilon}=2+\alpha_t/\nu_t=3$. More specifically, the diffusion coefficient of the energy-like density scales with the diverging correlation length $\xi$ as $D_\varepsilon \sim \xi^{-\alpha_t/\nu_t}$. 

The limiting case of an $N=2$-component order parameter corresponds to the tricritical point in $\mathrm{^3He}$--$\mathrm{^4He}$ mixtures, which is described by Model~F' \cite{PhysRevB.15.1427,Folk_2006}. Model~F' supplements Model~F of the Halperin-Hohenberg classification \cite{Hohenberg:1977ym} with an additional conserved density, corresponding to the concentration of $\mathrm{^3He}$. We have discussed that the universal critical dynamics of Model F'  is the same as the one of our (SSS') model for $N=2$. A first-order $\epsilon$-expansion around $d=4$ has suggested a weak-scaling behavior in Model~F' \cite{PhysRevB.15.1427}. While this is reproduced by our non-perturbative flow equations for small $\epsilon = 4-d$, beyond the perturbative $\epsilon$-expansion, we find that the strong-scaling behavior appears for $d \leq 3$, in particular also in $d\to 3^-$ dimensions as a limiting case, however. 

By adapting the original mode-coupling argument for the weak-scaling relation of the kinetic coefficients in Model~H \cite{Hohenberg:1977ym,Son:2004iv} to the tricritical point, we have argued that our results are not affected by a fluctuating momentum density of the fluid in $d=3$ spatial dimensions.

In QCD, non-vanishing current quark masses explicitly break the $O(4)$ chiral symmetry, turning the tricritical point into an ordinary $Z_2$ critical point. Its dynamic universality class is conjectured to be that of Model~H, i.e.,~that of the liquid-gas critical point \cite{Son:2004iv}. In Sec.~\ref{sec:explSymmBreak}, we have introduced an external field $H$ to explicitly break the $O(N)$ symmetry, and studied the eigenmodes of the linearized equations of motion around the mean-field solution. In the vicinity of the $Z_2$ line, we have found that for times of order $\xi_t\sim H^{-3/5}$ only one diffusive mode remains. This mode emerges from the mixing of  the (chiral) order parameter and the energy-like density, from which it inherits its diffusive nature as in the dynamic universality class of Model~B \cite{Son:2004iv}. Including the fluctuating momentum density is then expected to further lead to the dynamic universality class of Model~H, as described at the end of Sec.~\ref{sec:VB} above.

As also discussed in the main text, above the upper critical dimension of three the $\phi^6$ coupling becomes dangerously irrelevant, leading to a failure of hyperscaling relations. In our present framework, this is not properly incorporated. While this is not an issue for $d=3$, it would be interesting to investigate how dangerously irrelevant couplings affect the extraction of critical exponents from FRG flows, since it would allow us to extend the results of this work to $d>3$.

Moreover, it would be interesting to further study 
the tricritical point in $\mathrm{^3He}$--$\mathrm{^4He}$ mixtures, in the future. Most prominently, our finding of strong scaling at the tricritical point of the SSS' model for $N=2$ and $d=3$ was based on a non-perturbative truncation of the rFRG flow.
While the truncation is arguably well motivated by including all RG relevant couplings, it will be important to 
assess the remaining systematic uncertainties of the truncation,
and to verify our finding of strong scaling with ab initio methods such as numerical simulations of Model~F', e.g.~by extending recent simulation of Model~F~\cite{Chattopadhyay:2026dyd}.

Another motivation is the search for the QCD critical point
in heavy-ion collisions. Assuming proximity of the tricritical point for sufficiently small current quark masses, it would be interesting to study real-time dynamics and non-equilibrium phase transitions in the SSS' model for $N=4$.  For instance, non-equilibrium quenches within the four-component generalization of Model~G (i.e.,~the SSS model for $N=4$) were recently performed in Refs.~\cite{Florio:2025zqv,Florio:2025lvu}. These quenches show an enhancement in the infrared spectra of Goldstone modes, which would be a natural explanation for the observed excess of soft pions in heavy-ion collisions \cite{Devetak:2019lsk,Lu:2024shm}. In this spirit, it would be interesting to explore the phenomenological implications of quenches near the tricritical point of the SSS' model.

\acknowledgments
This work was supported by the Deutsche Forschungsgemeinschaft (DFG, German Research Foundation) through the CRC-TR 211 ‘Strong-interaction matter under extreme conditions’ -- project number 315477589 - TRR 211. MP is supported via Ramanujan Fellowship, Project File Number RJF/2025/000614.

\appendix

\section{FRG flow of effective potential \& exact solution at Gaussian fixed point}
\label{sec:exactSolEffPotGaussFP}

First of all, note that by setting the energy-like density $\varepsilon = \varepsilon_{\text{min},k}(\rho)$ to its $\rho$-dependent minimum, which satisfies
\begin{equation}
    \frac{\partial U_k}{\partial \varepsilon} \bigg\rvert_{\varepsilon_{\text{min},k}(\rho)} = 0 \,, \label{eq:eMin}
\end{equation}
we obtain the effective potential $V_k(\rho) \equiv U_k(\rho,\varepsilon_{\text{min},k}(\rho))$ of the order parameter,
which then satisfies the well-known flow equation 
\begin{align}
    \partial_k V_k = K_d k^{d+1} T \left( \frac{1}{2V_k' + 4\rho V_k''+k^2} + \frac{N-1}{2V_k'+k^2} \right)\, , \label{eq:flowEffPotRho}
\end{align}
as can be shown from \eqref{eq:flowEffPot}. In the truncation \eqref{eq:effPotTrunc} the scale-dependent effective potential $V_k(\rho)$ explicitly reads
\begin{align}
    V_k(\rho) = \frac{m_k^2}{2}\rho + \frac{\lambda'_k}{4!N} \rho^2 + \frac{\kappa_k}{6!N^2} \rho^3\, .
\end{align}
Expanding \eqref{eq:flowEffPotRho} around $\rho=0$ yields the standard flow equations \eqref{eq:m2LambdaKappaFlow} for $m_k^2$, $\lambda'_k = \lambda_k - 3\iota_k^2 \chi_{\varepsilon,k}$ 
and $\kappa_k$. 

The flow equation for the minimum $\varepsilon_{\text{min},k}(\rho)$ can be derived by requiring \eqref{eq:eMin} to hold at all FRG scales~$k$, which yields
\begin{align}
    \partial_k \varepsilon_{\text{min},k}(\rho) = -\frac{1}{\ddot{U}_k(\rho,\varepsilon)} \frac{\partial}{\partial \varepsilon}\partial_k U_k(\rho,\varepsilon) \bigg\rvert_{\varepsilon_{\text{min},k}(\rho)} . \label{eq:flowEqMin}
\end{align}
More generally, this suggests a $\rho$-dependent Taylor expansion of $U_k(\rho,\varepsilon)$ in $\varepsilon$ around $\varepsilon=\varepsilon_{\text{min},k}(\rho)$,
\begin{align}
    U_k(\rho,\varepsilon) = V_k(\rho) + \sum_{n=2}^{\infty} \frac{V_{n,k}(\rho)}{n!} (\varepsilon-\varepsilon_{\text{min},k}(\rho))^n \, ,\label{eq:UkExp}
\end{align}
with the series starting at $n=2$ due to \eqref{eq:eMin}. The coefficients $V_{n,k}(\rho)$ are given by
\begin{align}
    V_{n,k}(\rho) = \frac{\partial^n U_k}{\partial \varepsilon^n} \bigg\rvert_{\varepsilon_{\text{min},k}(\rho)} \,.
\end{align}
Their FRG flow is given by
\begin{align}
    \partial_k V_{n,k}(\rho)
    &= \frac{\partial^n}{\partial \varepsilon^n} \partial_k U_k \bigg\rvert_{\varepsilon_{\text{min},k}(\rho)} \hspace{-0.7cm} + \hspace{0.4cm} \label{eq:flowOfVn}  \frac{\partial^{n+1} U_k}{\partial \varepsilon^{n+1}} \bigg\rvert_{\varepsilon_{\text{min},k}(\rho)} \hspace{-0.7cm} \partial_k \varepsilon_{\text{min},k}(\rho) \,.
\end{align}
This method can be used to derive flow equations for the functions $V_{n,k}(\rho)$ for arbitrary $n$, in principle. In the main text, we employ the truncation \eqref{eq:effPotTrunc} for the effective potential $U_k(\rho,\varepsilon)$. We use the following projections to obtain $\chi_{\varepsilon,k}$ and $\iota_{k}$ from  $U_k(\rho,\varepsilon)$,
\begin{align}
    \frac{1}{\chi_{\varepsilon,k}} &= \frac{\partial^2 U_k}{\partial \varepsilon^2} \bigg\rvert_{\rho=0,\varepsilon=\varepsilon_0} \hspace{-1.0cm} = V_{2,k}(0) \,, \\
    \iota_{k} &= \frac{\delta_{ab}}{\sqrt{N}}
    \frac{\partial^3 U_k}{\partial\phi_a \partial \phi_b \partial \varepsilon} \bigg\rvert_{\rho=0,\varepsilon=\varepsilon_0} \hspace{-1.0cm} = -2\sqrt{N} V_{2,k}(0) \varepsilon_{\text{min},k}(0) \,.
\end{align}
Using \eqref{eq:flowEqMin} and \eqref{eq:flowOfVn} with $n=2$ to derive flow equations for $\varepsilon_{\text{min},k}(0)$ and $V_{2,k}(0)$, respectively, and inserting the truncation \eqref{eq:effPotTrunc} of the effective potential, one arrives at the anticipated flow equations \eqref{eq:iotaFlow} and 
\eqref{eq:chiFlow} for the scale-dependent  coupling $\iota_{k}$ to the energy-like density $\varepsilon$ and its 
susceptibility $\chi_{\varepsilon,k}$.

\newcommand{\newVar}{\zeta}

We now proceed to find an analytic solution of the $k$-scaling of the effective potential $U_k(\rho,\varepsilon)$ at the Gaussian fixed point. First of all, note that at the Gaussian fixed point $V_k(\rho)$ is a $k$-dependent constant corresponding to the (irrelevant) ground-state energy density. We use the following ansatz for the full effective potential $U_k(\rho,\varepsilon)$,
\begin{align}
    U_k(\rho,\varepsilon) &= W_k(\varepsilon - \varepsilon_{\text{min},k}(\rho)) \, , \quad\text{with}  \nonumber \\ \varepsilon_{\text{min},k}(\rho) &= \varepsilon_{0,k} - \frac{\iota_k \chi_{\varepsilon,k}}{2\sqrt{N}}\rho \,, \label{eq:formOfEffPot}
\end{align}
where $W_k(\newVar)$ denotes some yet undetermined function that depends on $\newVar \equiv \varepsilon - \varepsilon_{\text{min},k}(\rho)$.  In particular, this ansatz implies that the coefficients
\begin{equation}
V_{n,k}(\rho) = \frac{\partial^n W_k}{\partial \newVar^n} \bigg\rvert_{\newVar=0}
\end{equation}
in \eqref{eq:UkExp} are $\rho$-independent. 
Eq.~\eqref{eq:eMin} entails $\dot{W}_k(0) = 0$ (where we consistently use the dot to denote a derivative with respect to $\zeta$). Eq.~\eqref{eq:flowEqMin} yields flow equations for the coefficients in (\ref{eq:formOfEffPot}),
\begin{align}
    \partial_k \varepsilon_{0,k} &= K_d k^{d-3} T\sqrt{N}  \iota_k \chi_{\varepsilon,k} \,, \nonumber \\ \partial_k \left(- \frac{\iota_k \chi_{\varepsilon,k}}{2\sqrt{N}}\right) &= 0 \, ,  \label{eq:abFlow} 
\end{align}
which entails $\iota_k \chi_{\varepsilon,k} = const$. We denote this constant by $b \equiv \iota_k \chi_{\varepsilon,k}$.
Inserting \eqref{eq:formOfEffPot} into \eqref{eq:flowEffPot} and using \eqref{eq:abFlow} yields a closed flow equation for $W_k(\newVar)$,
\begin{align}
    \partial_k W_k(\newVar) = -2 K_d k^{d-3} TN b \dot{W}_k(\newVar) + \frac{K_d k^{d+1} TN}{k^2-2b \dot{W}_k(\newVar)} \, . \label{eq:flowEqWk}
\end{align}
Differentiating with respect to $\newVar$ reveals an advection equation for $w_k(\newVar) \equiv \dot{W}_k(\newVar)$,
\begin{align}
    \partial_k w_k = -\partial_{\newVar}\left\{ K_d k^{d-3} TN \left( 2 b w_k - \frac{k^4}{k^2-2b w_k} \right) \right\} \,.
\end{align}
To study fixed-point solutions, we introduce the dimensionless variables $\bar{\newVar} \equiv k^{2-d}\newVar/bNT$ and $\bar{w} \equiv k^{-2} b w_{k}$, for which the flow equation becomes
\begin{align}
    k\partial_k \bar{w} = -2\bar{w} + (d-2)\bar{\newVar}\bar{w}' - \partial_{\bar{\newVar}} \left\{ K_d \left( 2
    \bar{w}-\frac{1}{1-2
    \bar{w}}\right) \right\}\, ,
\end{align}
with $\dot{\bar{w}} \equiv \partial\bar{w}/\partial\bar{\zeta}$.
At the fixed point we have $k\partial_k \bar{w} = 0$, which yields a first-order ordinary differential equation for $\bar{w}$,
\begin{align}
    0 = -2\bar{w} + (d-2)\bar{\newVar}\dot{\bar{w}} - 2
    K_d \left( 1-\frac{1}{(1-2
    \bar{w})^2}\right) \dot{\bar{w}} \,.\label{eq:wBarODE}
\end{align}
The inverse susceptibility $1/\chi_{\varepsilon} = \ddot{U}_k(0,0) = \ddot{W}_k(0)$ of the energy-like density is related to the fixed-point potential via
\begin{equation} \label{Eq:chieAtGaussian}
    \frac{1}{\chi_{\varepsilon}} = \ddot{W}_k(0) = \dot{w}_k(0) = \frac{k^{4-d}}{b^2NT} \dot{\bar{w}}(0) \,.
\end{equation}
An expansion of \eqref{eq:wBarODE} around $\bar{\newVar}=0$ yields the Taylor coefficients
\begin{align}
    \bar{w}(0) = 0\,, \quad \dot{\bar{w}}(0) = \frac{4-d}{8 K_d} \,, \quad \ldots 
\end{align}
which entails that the inverse susceptibility behaves as $\chi_{\varepsilon} \sim k^{-(4-d)}$, implying $\alpha_t/\nu_t = 4-d$, as expected at the Gaussian fixed point. The dimensionless coupling $\bar{v} = k^{d-4} T \iota_{k}^2 \chi_{\varepsilon,k}$ from the main text is related to the fixed-point potential via (using $\iota_{k} = 2\sqrt{N} \dot{U}'_k(0,0) = -2\sqrt{N} b \ddot{W}_k(0)$)
\begin{align}
    \bar{v} 
    &= 4\dot{\bar{w}}(0) 
    = \frac{4-d}{2K_d} \, ,
\end{align}
in agreement with our result \eqref{eq:GaussFP} in the main text.

\section{\boldmath{$x_{\Gamma^{\phi}}$} inequality}
\label{app:ineq}

In this appendix, we show that $x_{\Gamma^{\phi}}$ is bounded from below. We assume $2<d<4$ and $N \geq 1$. As in the main text, we split $x_{\Gamma^{\phi}} = x_{\Gamma^{\phi}}^{(n)} + x_{\Gamma^{\phi}}^{(\varepsilon)}$ into contributions from fluctuations of the $O(N)$ charge densities and from fluctuations of the energy-like density, which are given in \eqref{eq:xGammaPhiN} and \eqref{eq:xGammaPhiE}, respectively.

The contribution $x_{\Gamma^{\phi}}^{(n)}$ is non-negative: At the Gaussian fixed point \eqref{eq:GaussFP}, $x_{\Gamma^{\phi}}^{(n)}$ can be written as
\begin{align}
    x_{\Gamma^{\phi}}^{(n)} &= \frac{(N-1)f}{d-2} \bigg( \frac{2+(4-d)
   w_n}{w_n+1} \;- \\ \nonumber &\hspace{2.5cm} (4-d) \, _2F_1\left(1,\frac{d}{2}-1;\frac{d}{2};-\frac{1}{w_n}\right)\bigg) \, .
\end{align}
Using the Euler integral representation of the hypergeometric function (assuming $0< w_n<\infty$), we find the inequality 
\begin{align}
    {}_2F_1\left(1,\frac{d}{2}-1;\frac{d}{2};-\frac{1}{w_n}\right) &=  \frac{d-2}{2}\int_0^1 \frac{dx \,x^{d/2-2}}{1+x/w_n}  \nonumber \\ 
    &< \frac{d-2}{2}\int_0^1 dx \,x^{d/2-2} \nonumber  \\ &= 1 \,,
\end{align}
which implies
\begin{align}
    x_{\Gamma^{\phi}}^{(n)} &= \frac{(N-1)f}{d-2} \bigg( \underbrace{\frac{2+(4-d)
   w_n}{w_n+1}}_{>4-d} \;- \nonumber \\  &\hspace{2.5cm} (4-d) \, \underbrace{_2F_1\left(1,\frac{d}{2}-1;\frac{d}{2};-\frac{1}{w_n}\right)}_{<1}\bigg) \nonumber \\
   &>\frac{(N-1)f}{d-2} \geq 0 \,.
\end{align}

On the other hand, the contribution
$x_{\Gamma^{\phi}}^{(\varepsilon)}$ can be negative but is bounded from below: At the Gaussian fixed point \eqref{eq:GaussFP}, $x_{\Gamma^{\phi}}^{(\varepsilon)}$ can be written as
\begin{align}
    &x_{\Gamma^{\phi}}^{(\varepsilon)} = \\ \nonumber
    &-\frac{4-d}{N} \bigg( \frac{dw_{\varepsilon}}{2(1+w_{\varepsilon})} + \frac{4-d}{2} \, {}_2F_1\left(1,\frac{d}{2};\frac{d}{2}+1;-\frac{1}{w_{\varepsilon }}\right)\bigg)\, .
\end{align}
Using the Euler integral representation of the hypergeometric function (again assuming $0 < w_{\varepsilon}<\infty$), we find
\begin{align}
    {}_2F_1\left(1,\frac{d}{2};\frac{d}{2}+1;-\frac{1}{w_{\varepsilon }}\right) &=  \frac{d}{2}\int_0^1 \frac{dx \,x^{d/2-1}}{1+x/w_{\varepsilon}}  \nonumber \\ 
    &< \frac{d}{2}\int_0^1 dx \,x^{d/2-1} \nonumber  \\ &= 1 \,,
\end{align}
which yields
\begin{align}
    &x_{\Gamma^{\phi}}^{(\varepsilon)} = \nonumber \\ 
    &-\frac{4-d}{N} \bigg( \underbrace{ \frac{dw_{\varepsilon}}{2(1+w_{\varepsilon})} }_{< d/2} +
    \frac{4-d}{2} \, \underbrace{ {}_2F_1\left(1,\frac{d}{2};\frac{d}{2}+1;-\frac{1}{w_{\varepsilon }}\right)}_{< 1} \bigg) \nonumber \\
    &\phantom{x_{\Gamma^{\phi}}^{(\varepsilon)}} > -\frac{2(4-d)}{N} \,.
\end{align}
Together with $x_{\Gamma^{\phi}}^{(n)} \geq 0$ we obtain the anticipated result,
\begin{align}
    x_{\Gamma^{\phi}} > -\frac{2(4-d)}{N} \,.
\end{align}

\section{Specific heat near the tri-critical point}
\label{App:cv}

The universal form of the singular part of the pressure near a tricritical point can be expressed in the following scaling form \cite{Hatta:2002sj}:
\begin{eqnarray}
P(\mu_B,T)=A r^{2-\alpha_t} g(h\,r^{-\beta_t\delta_t},\lambda\,r^{-\phi_t})\, ,
\end{eqnarray}
where $A$ is a constant of suitable dimension and $r\equiv m^{2}$. In case of QCD, the ordering field $h$ is proportional to quark mass $m_q$. 
It suffices to use a linear dependence of $r$ and $\lambda$ on $\mu_B$ and $T$ to describe the leading singular behavior:
\begin{eqnarray}
r&=&r_{\mu_B} (\mu_B-\mu_{B\text{tcp}})+r_{T}(T-T_{\text{tcp}})\, ,\\
\lambda&=&\lambda_{\mu_B} (\mu_B-\mu_{B\text{tcp}})+\lambda_{T}(T-T_{\text{tcp}}) \, .
\end{eqnarray}

The tri-critical exponents $\alpha_t, \, \beta_t, \, \delta_t$ and $\phi_t$ are equal to $1/2, 1/4, 5$ and $1/2$, respectively. The fluctuations diverge most strongly along the $h$ direction, less strongly along the $r$ direction, and the singular contribution vanishes along the $\lambda$ direction. This can be seen from the following scaling relations:
\begin{eqnarray}\nonumber
P_{hh}&\sim &r^{2-\alpha_t-2\beta_t\delta_t} \sim r^{-1} \, , \quad P_{hr}\sim r^{1-\alpha_t-\beta_t\delta_t} \sim r^{-3/4} \, , \\ \nonumber P_{rr}&\sim& r^{-\alpha_t} \sim r^{-1/2}\, ,\quad P_{r\lambda}\sim r^{1-\alpha_t-\phi_t} \sim r^{0}\, , \\&\,& P_{\lambda\lambda}\sim r^{2-\alpha_t-2\phi_t} \sim r^{1/2} \, , 
\end{eqnarray}
where we have denoted subscripts to imply derivatives of $P$ with respect to the variable mentioned, while the other variables are kept constant. For example, $P_{rr}$ denotes $\partial^{2}P/\partial r^2$ taken along $h=\lambda=\text{const}$. The second derivatives of $P$ with respect to $T$ and $\mu_B$, denoted by $P_{TT}, P_{\mu_B\mu_B}$ and $P_{\mu_B T}$ diverge as $r^{-\alpha_t}$ near the tri-critical point in the chiral limit, as $T$ and $\mu_B$ do not couple linearly to $m_q$. 

The two commonly discussed second derivatives of pressure are the specific heats at constant pressure and constant volume, $c_p$ and $c_v$ respectively. Close to the line of $Z_2$ critical points, these two quantities diverge with different exponents, $c_p\sim (T-T_c(m_q))^{-\gamma}$ and $c_v\sim (T-T_c(m_q))^{-\alpha}$, with $\alpha >0 $ in the three-dimensional Ising universality class. However, in the chiral limit, near the tricritical point, $c_p\sim r^{-\alpha_t}$ but $c_v$ remains finite. The isobaric specific heat $c_p$ is given by: 
\begin{align}
    c_p&=T\left(\frac{\partial \hat{s}}{\partial T}\right)_{p} \nonumber \\
    &=\frac{T}{n_B}\left[\left(\frac{\partial s}{\partial T}\right)_{p}-\frac{s}{n_B}\left(\frac{\partial n_B}{\partial T}\right)_{p}\right] \nonumber  \\
    &=\frac{T}{n_B}\left[P_{TT}-2\hat{s}P_{\mu_B T}+\hat{s}^2P_{\mu_B\mu_B} \right] \nonumber \\
    &\overset{m_q \to 0}{=} \frac{T}{n_B} P_{rr}\left(r_T-\hat{s}\, r_{\mu_B}\right)^2+\dots \sim r^{-\alpha_t} \,.
\end{align}

The specific heat at constant volume, or equivalently, at constant number density, in the chiral limit is given by:
\begin{align}
c_v&=\frac{T}{n_B}\left(\frac{\partial s}{\partial T}\right)_{n} \nonumber \\ 
&=\frac{T}{n_B}\left(P_{TT}-\frac{(P_{T\mu_B})^2}{P_{\mu_B\mu_B}}\right) \nonumber  \\ 
&=\frac{T}{n_B P_{\mu_B\mu_B}}\left(P_{TT}P_{\mu_B\mu_B}-(P_{T\mu_B})^2\right) \nonumber \\ 
&= c^{\text{tcp}}_v+\frac{T}{n_B}\left(\frac{r_T}{r_{\mu_B}}\lambda_{\mu_B}-\lambda_T\right)^2\left[P_{\lambda\lambda}-\frac{P^2_{r\lambda}}{P_{rr}}\right]+O(r) \nonumber  \\
&= c^{\text{tcp}}_v+O(|r|^{1/2})\, ,
\label{eq:cv}
    \end{align}
    where $c_v^{\text{tcp}}$ denotes the finite value of the isochoric specific heat at  the tricritical point. 
Here, again we have used subscripts to imply derivatives. When the subscript $T$ is written the derivative is taken with respect to $T$ while keeping $\mu_B$ constant. The combination $\left(P_{TT}P_{\mu_B\mu_B}-(P_{T\mu_B})^2\right)$ subtracts off the dominant singular fluctuations along the $Z_2$ order parameter direction, thereby projecting out the remaining singular component which is subleading. At the tricritical point, the contour of the number density becomes parallel to the $\lambda$ line.

\begin{figure}[t]
    \centering
    \includegraphics[width=\linewidth]{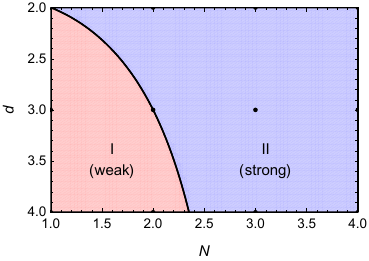}
    \caption{Transition line that separates weak-scaling and strong-scaling behavior at the Gaussian fixed point in the $(N,d)$-plane, given by Eq.~\eqref{eq:NSS}. Integer values of $d$ and $N$ are indicated by dots. See text for an important caveat concerning the applicability of the Gaussian fixed point.}
    \label{fig:pd}
\end{figure}

We shall evaluate $P_{rr}$  in mean-field theory, below. In mean field, the pressure near the tricritical point can be expressed as the minimum of a $\phi^{6}$ potential as follows:
\begin{eqnarray}
P(\mu_B,T)=-A\, \min_{\phi} V(\phi; r,\lambda)\,,
\end{eqnarray}
where
\begin{eqnarray}
V(\phi;r,\lambda)=\frac{r}{2}\phi^{2} +\frac{\lambda}{4!}\phi^{4}+\phi^{6} -h \phi\,.
\end{eqnarray}
$P_{rr}$ can be calculated as:
\begin{eqnarray}
P_{rr}=-A \bar{\phi}\frac{\partial\bar{\phi}}{\partial r} =A\left.\frac{(V_{r\phi})^2}{V_{\phi\phi}}\right|_{\bar{\phi}},
\end{eqnarray}
where $\bar{\phi}$ is the value of $\phi$ which minimizes $V(\phi;r,\lambda)$.
Similarly, 
\begin{eqnarray}
P_{r\lambda}=A \frac{V_{r\phi}V_{\lambda\phi}}{V_{\phi\phi}}\, , \quad P_{\lambda\lambda}=A \frac{(V_{\lambda\phi})^2}{V_{\phi\phi}}\, .
\end{eqnarray}

In the symmetry restored phase $\bar{\phi}=0$, therefore, the singular contribution to $P_{rr}=0$ in this phase. When $\bar{\phi}\neq0$, the singular contribution to $P_{rr}$ in mean field is given by:
\begin{eqnarray}
P_{rr}=\frac{3A}{\sqrt{-864\,r+\lambda^{2}}}\,.
\end{eqnarray}
One can see that away from the tricritical point, since $\lambda^2\gg 0$,
\begin{eqnarray}
    P_{rr}\propto \lambda^{-1}
\end{eqnarray}
is finite along the $\lambda$ line and close to the tri-critical point:
\begin{eqnarray}
    P_{rr}\propto r^{-1/2} \propto (T-T_{\text{tcp}})^{-1/2},
\end{eqnarray}
where $T_{\text{tcp}}$ is the temperature at the tricritical point. 

\bigskip

\section{Phase diagram in \boldmath{$(N,d)$}-plane}
\label{sec:pd}

Fig.~\ref{fig:FPMergerN2} in the main text shows that for $N=2$ there is a transition from strong-scaling to weak-scaling behavior at $d=3$. It is instructive to map out the transition line that separates strong-scaling and weak-scaling behavior in the full $(N,d)$ plane. Strong scaling occurs if and only if \eqref{eq:wnFPEq} has a non-zero solution $0< w_n^* < \infty$. Such a solution exists if $N \geq N_{l}(d)$ with
\begin{equation}
    N_{l}(d) = 1-\frac{1}{d}+\frac{\sqrt{5d^2-10d+1}}{d} \,. \label{eq:NSS}
\end{equation}
Fig.~\ref{fig:pd} shows the resulting transition line in the $(N,d)$ plane. The line $N=2$ corresponds to Fig.~\ref{fig:FPMergerN2} from the main text, where we see that the transition happens precisely at $d=3$. Fig.~\ref{fig:pd} extends this result to all $N \geq 1$. Qualitatively, we see that as $N$ is increased the dimensionality required for strong scaling also increases. We observe that weak-scaling behavior is generally restricted to small values of $N$. In particular, we find that for $d \leq 4$ and $N \geq N_l(4) = (4+\sqrt{41})/3 \approx 2.35$, there is no weak-scaling region anymore. 

The same important caveat that we have discussed towards the end of Sec.~\ref{sec:resultsDynamics} applies: We have assumed that $(i)$ tricritical behavior is governed by the Gaussian fixed point which is strictly true only for $d\ge 3$, and $(ii)$ that there are no dangerously irrelevant couplings which is guaranteed only for $d\le 3$. These assumptions are simultaneously satisfied only at the upper critical dimension $d=3$, where Fig.~\ref{fig:pd} reflects our result from Sec.~\ref{sec:resultsDynamics} that we find strong scaling for all $N\geq 2$. The physically interesting case of the tricritical point in $\mathrm{^3He}$--$\mathrm{^4He}$ mixtures with $N=2$ in $d=3 $ dimensions is precisely on the separation line which explains why strong dynamic scaling is observed although the fixed point value of the ratio $w_n$ of relaxation times vanishes there, thus apparently only applying weak dynamic scaling at the tricritical point in $\mathrm{^3He}$--$\mathrm{^4He}$ mixtures which are thus special in several ways. 

\begin{widetext}

\section{1PI vertex functions for kinetic coefficients}
\label{sec:1PIVertFnc}

The relevant `classical' vertices (i.e.~those with one response field) are given by:
\begin{align}
    \Gamma_k^{\phi_a \phi_b \tilde{N}_{cd}}(\Dvec{p},\Dvec{q},\Dvec{r}) &=g \frac{i (\omega_q-\omega_p)}{\Gamma^{\phi}_{k}\,\gamma_{k}\vec r^2} ~  (\delta_{ac}\delta_{bd} - \delta_{ad}\delta_{bc}) ~ (2\pi)^{d+1} \delta(\Dvec{p}+\Dvec{q}+\Dvec{r}) \, ,
\end{align}
\begin{align}
    \Gamma_k^{\phi_a \tilde{\Phi}_b n_{cd}}(\Dvec{p},\Dvec{q},\Dvec{r}) &= -g \frac{i\omega_r}{\Gamma^{\phi}_{k}\,\gamma_{k}\vec r^2} ~ (\delta_{ac}\delta_{bd} - \delta_{ad}\delta_{bc}) ~ (2\pi)^{d+1} \delta(\Dvec{p}+\Dvec{q}+\Dvec{r}) \, ,
\end{align}
\begin{align}
    \Gamma_k^{n_{ab} n_{cd} \tilde{N}_{ef}}(\Dvec{p},\Dvec{q},\Dvec{r}) =   \frac{g}{\,\gamma_{k}\vec r^2} \bigg[ &\frac{i \omega_p}{\gamma_{k}\vec p^2} (  \delta_{ae}\delta_{bc}\delta_{df} - \delta_{af}\delta_{bc}\delta_{de} - \delta_{ac}\delta_{be}\delta_{df} + \delta_{ac}\delta_{bf}\delta_{de}  \nonumber \\
    &\phantom{\frac{iZ_{n,k}^\omega p^0}{\gamma_{n,k}(\vec p)} (} \hspace{-0.5cm} - \delta_{ae}\delta_{bd}\delta_{cf} + \delta_{af}\delta_{bd}\delta_{ce} + \delta_{ad}\delta_{be}\delta_{cf} - \delta_{ad}\delta_{bf}\delta_{ce} )  \nonumber \\
    +&\frac{i\omega_q}{\gamma_{k}\vec q^2} (  \delta_{ad}\delta_{bf}\delta_{ce} - \delta_{ad}\delta_{be}\delta_{cf} - \delta_{ac}\delta_{bf}\delta_{de} + \delta_{ac}\delta_{be}\delta_{df} \nonumber \\
    &\phantom{\frac{iZ_{n,k}^\omega q^0}{\gamma_{n,k}(\vec q)} (} \hspace{-0.5cm} - \delta_{bd}\delta_{af}\delta_{ce} + \delta_{bd}\delta_{ae}\delta_{cf} + \delta_{bc}\delta_{af}\delta_{de} -\delta_{bc}\delta_{ae}\delta_{df} ) \bigg] \times \nonumber \\
    &(2\pi)^{d+1} \delta(\Dvec{p}+\Dvec{q}+\Dvec{r}) \, .
\end{align}
The relevant `anomalous' vertices (i.e.,~those that involve two response fields) are given by:
\begin{align}
    &\Gamma_k^{\phi_a \phi_b \tilde{\Phi}_c \tilde{\Phi}_d}(\Dvec{p},\Dvec{q},\Dvec{r},\Dvec{s}) =   \frac{2g^2 iT}{(\Gamma^{\phi}_{k})^2} \times \nonumber \\
    &\hspace{0.8cm} \bigg[ \delta_{ac}\delta_{bd} \frac{1}{\gamma_{k}(\vec p+\vec r)^2} + \delta_{ad}\delta_{bc} \frac{1}{\gamma_{k}(\vec p+\vec s)^2}
     - \delta_{ab}\delta_{cd}\left( \frac{1}{\gamma_{k}(\vec p+\vec r)^2} + \frac{1}{\gamma_{k}(\vec p+\vec s)^2}  \right) \bigg] \times \nonumber \\
     &\hspace{1.0cm} (2\pi)^{d+1} \delta(\Dvec{p}+\Dvec{q}+\Dvec{r}+\Dvec{s})  \, ,
\end{align}
\begin{align}
    \Gamma_k^{\phi_a \phi_b \tilde{N}_{cd} \tilde{N}_{ef}}(\Dvec{p},\Dvec{q},\Dvec{r},\Dvec{s}) &=  \frac{2 g^2 iT}{\Gamma^{\phi}_{k}\,\gamma_{k}\vec r^2\,\gamma_{k}\vec s^2} \times \nonumber \\  
    &\hspace{0.8cm} \Big[  + \delta_{ad}\delta_{be}\delta_{cf}+\delta_{bd}\delta_{ae}\delta_{cf} - \delta_{ad}\delta_{ce}\delta_{bf} - \delta_{bd}\delta_{ce}\delta_{af} \nonumber \\  
    &\hspace{1.0cm} - \delta_{ac}\delta_{be}\delta_{df} - \delta_{bc}\delta_{ae}\delta_{df} + \delta_{ac}\delta_{de}\delta_{bf} + \delta_{bc}\delta_{de}\delta_{af} \Big] \times  \nonumber \\
    &\hspace{1.0cm} (2\pi)^{d+1} \delta(\Dvec{p}+\Dvec{q}+\Dvec{r}+\Dvec{s}) \, ,
\end{align}
with $\Dvec{p} = (\omega_p,\vec{p})^T$, $\Dvec{q} = (\omega_q,\vec{q})^T$, $\Dvec{r} = (\omega_r,\vec{r})^T$ and $\Dvec{s} = (\omega_s,\vec{s})^T$.

\end{widetext}

\bibliographystyle{h-physrev3}
\bibliography{refs}

\end{document}